\documentclass[12pt,preprint]{aastex}

\slugcomment{\today}

\usepackage{epsfig}
\usepackage{times}
\usepackage{xcolor}
\usepackage{graphicx}

\shorttitle{Perturbed Torus by Kicked Black Hole}
\shortauthors{Donmez}

\begin{document}


\title{On the dynamics of the torus around the kicked black hole}


\author{O. Donmez\altaffilmark{1}, Anwar Al-Kandari\altaffilmark{1}, and Ahlam Abu Seedou\altaffilmark{1}} 
\affil{College of Engineering and Technology, American
  University of the Middle East, Kuwait}

\altaffiltext{1}{College of Engineering and Technology, American
  University of the Middle East, Kuwait}


\begin{abstract}
  There is a special interest to understand the dynamical properties of the accretion disk created
  around the newly formed black hole
  due to the supermassive black hole binaries which merge inside the gaseous disk. The newly formed black hole
  would have a kick velocity at up to thousands of km/s that drives a
    perturbation on a newly accreted torus around the black hole. Some of the observed
    supermassive black holes at the center
    of the Active Galactic Nucleus (AGN) moves with a certain velocity relative to its broader accretion disk.
    In this paper, the effects of the kicked black holes onto the infinitesimally thin accreted 
    torus are studied  by using
  the general relativistic hydrodynamical code, focusing on changing the dynamics of the accretion disk during the
  accretion disk-black hole interaction. We have  found the non-axisymmetric global mode  $m=1$
  inhomogeneity, which causes a spiral-wave-structure, is 
  excited on the torus due to kicked black hole.
  The higher the perturbation velocity produced by the kicked black hole,
  the longer time the torus takes to reach the saturation point. The created spiral density waves which rapidly
  evolve into the spiral shocks are also
  observed from the numerical simulations. The spiral shock  is responsible for accreting matter
  toward  the black hole.
  Firstly, the spiral-wave-structure
    is developed and the accretion through the spiral arms  is stopped
  around the black hole. At the later time of simulation, the formed spiral shocks  partly
  causes the angular momentum loss across the torus.

\end{abstract}

\keywords{general relativistic hydrodynamics:  numerical relativity: 
spiral wave: torus: instability: kicked black hole}


\section{Introduction}
\label{Introduction}

In recent years, number of numerical simulations
\citep{Baker1, KorLov, Gerosa1} and
astronomical observations \citep{Webb_other1, Bustillo_other1, Shaughnessy1, Atri1}
have revealed the kicked black hole at the center of the black hole X-ray binaries
and the AGN. This has occurred due to the asymmetric momentum loss of the black hole,
either in a case of black hole mergers or in the release of the gravitation radiation
during this process.
It was found from the numerical simulation that the gravitational wave recoil velocities range
  from the smallest possible value up to  thousands of $km/s$ \citep{Lousto1, Ponce1}.
  The supermassive black hole kick velocities were also  measured  from the smallest possible value up to
  $500 km/s$ \citep{Gerosa2}. It is also expected to be a kicked velocity from the binary coalescence in the galaxy
  mergers. Kick velocities up to a few hundred $km/s$ of the stellar mass black holes formed in the
  core-collapse supernovae were first described by \citet{Bekenstein1}  which might account for
  their relatively small branching ratio to long GRBs\citet{Putten1}.

A smoothed particle hydrodynamics numerical simulation of
the impulsive kick transmitted to a black hole on the dynamical evolution of the accretion disk has been
done by \citet{Ponce1}. They have found the higher potential luminosity considering the region $0.4pc$
from the galactic center for more oblique kick angle.
The occurrence of the kicked black hole inside the gaseous disks perturbs the surrounding gas
and it could create a signature of the electromagnetic radiation which gives an evidence of merging.
The kick directed into the equatorial plane was studied by \citet {Corrales1} using the $2D$ FLASH code.
The one-armed spiral shock wave appeared as a characteristic respond and it produced  a total luminosity up to
$10\%$ of the Eddington luminosity. The generated shocks' energy can produce electromagnetic flares that
could last up to hundreds of thousands years \citep{Shields1, Schnittman1}.

One of the new observations was made in a distant galaxy which is known as
quasar $3C$ $186$ \citep{Chiaberge1}.
It is the most massive black hole which is kicked out
of its central home. 
The merging two black holes produces gravitational waves
which carry energy and angular momentum of the system.
Finally, when they merge in a violent collision, and the released energy is enough,
the newly formed black hole would be kicked
away from the center of the galaxy or even out of the entire galaxy in the opposite side of the location
with the highest gravitational waves.
The range of the kick can  vary depending on the properties of these two colliding black holes,
their separation, and inclination angles.

The more observations showed that the broad-line emission system has a different red-shift
when it is compared with the narrow-line emission system in several AGNs, {Chiaberge1}.
such as  $SDSSJ092712.65+294344.0$ with a blueshift of $2650 km/s$ \citep{KomZh},
  $SDSS$ $J105041.35+345631.3$ with a kicked velocity  $3500 km/s$ \citep{Shields2}, $CXOC$ $J100043.1+020637$
  with a kicked velocity $1200 km/s$ \citep{Civano1}, and $E1821+643$ with a kicked velocity $2100 km/s$
 \citep{Rabinson1}. 
 When the black hole is kicked, broad-line emitting region is dragged out along with the kicked black hole
while the narrow-line region is left behind \citep{Chiaberge1}.
It is believed that Radio-loud AGN with strong relativistic jets could contain the rapidly spinning
black holes.
Fastly spinning black holes could lead the accreted matter getting close to the
  black hole horizon and create a strong interaction between the matter and black hole.

High energetic astrophysical sources such as Gamma Ray Bursts (GRBs) \citep{KumarZhang},
X-ray binaries \citep{RemMcC1},
and AGNs \citep{Reynolds1} are powered by
whether the accretion mechanism onto the black holes or
  the spin energy of the black hole \citep{Putten3}.
A high accretion rate could be presented during the development of the instability. 
 One of those instability is created by gravitational instability \citep{Rice1, Rice2}, which is
prone the development of
the non-axisymmetric instability, called spiral density waves.
The time dependent rest-mass density causes the material to concentrate at certain places
  in the effective potential around the black hole and generates spiral arms.
  The arms slowly rotates  and
  allow the the disk angular momentum transform outwards \citep{Blaes1}.
  Spiral structure mechanism leads developing a global
  non-axisymmetric modes in torus. The accreted matter belongs to spiral arms
  are characterized by the high internal temperatures which cause a
  high-energy emission from such sources as AGNs, $X-$ray binary, and GRBs \citep{Reynolds1,RemMcC1}.
During the development of spiral structure, some vigorous dynamics of the torus would be observed.
The vigorous  phenomena happening on the torus dynamics around the kicked black hole
would cause the conversion of the gravitational binding energy into thermal and kinetic
energies during the process of the losing the angular momentum. 

The oscillating properties of tori around the black hole and  neutron stars have been
investigated theoretically and numerically by perturbing the
system \citep{RezYosMac, AbramFrag}. The numerical simulations of the radially
perturbed accretion torus around the rotating and non-rotating black hole are
performed by \citep{Lee1}. The perturbation of the black hole torus system can be observed
in different types of perturbations such as angular velocity perturbation \citep{Donmez3},
spherical shell accretion \citep{Donmez4}, radial velocity perturbation \citep{Donmez2},
and Bondi-Hoyle accretion \citep{Donmez5}, which are all studied
to extract the oscillation properties of the torus' PPI instabilities, and  the quasi-periodic
frequencies produced during these oscillations. The non-axisymmetric perturbations trigger
not only the PPI instability but also spiral waves
  and  cause the excitation of frequencies due to the non-linear coupling  of
the modes. Doing numerical simulation enables us to study the response of the accreted torus
if the kicked is directed into the equatorial plane of the accretion disk.

The first numerical results  of  perturbation of the initially stable torus due
to the  kicked black hole at the center of the system are reported. To this extent,
we solve the general relativistic
hydrodynamical equations in a fixed Kerr space-time metric to investigate how a torus reacts
to the kicked black hole.
The outline of the paper is as follows.
The computational framework of the problem with the physical variables of stable torus,
properties of the black hole, and the  radial velocities assumed created by kicked black holes
are given in Section \ref{Computational Framework of the 2D Model}. In Section \ref{Numerical Results},
dynamical properties of the kicked torus found from the numerical simulations are provided around the rotating
and the non-rotating black holes.
The spiral wave mode $m=1$ generated by gravitational instability
  is reported and
the dynamical structure of perturbed
torus around the rotating black hole with the non-rotating case is compared. Finally, the summary of our results
is given in Section \ref{Discussion and Conclusion}. Unless otherwise specified, throughout
this paper, the geometrized unit, $c = G =1$ has been used.
 

\section{Computational Framework of the 2D Model}
\label{Computational Framework of the 2D Model}


To model the effects of the kicked black hole on the stable torus, we solve the General Relativistic
Hydrodynamical (GRH) equations on the equatorial plane. GRH equations are described for the
relativistic perfect fluid in the spacetime coordinate which can be written in covariant and 
conservative forms.  The covariant form of GRH equation is

\begin{eqnarray}
    \bigtriangledown_{a}T^{ab}=0,  \;\;\;\;\;
  \bigtriangledown_{a}(\rho u^{a})=0,
\label{GRH1}
\end{eqnarray}

\noindent $T^{ab}  = \rho h u^{a}u^{b} + P g^{ab}$ stress-energy-momentum tensor for perfect
fluid.
The tensor depends on the specific enthalpy $h = 1 + \epsilon + \frac{P}{\rho}$,
  the rest mass density $\rho$, pressure $P$, the four-velocity  $u^{a}$, and the four--matrix
  of the space-tiem $g^{\mu \nu}$.
The perfect fluid equation of state
$P = (\Gamma - 1)\rho \epsilon$ is used to constrain the system. After using $3+1$ formalism,
Eq.\ref{GRH1} can be written conservative form. The formulation of GRH equation and its numerical
solution were discussed with the full details  in \citet{Donmez1} (see also
\citep{Donmez6, Donmez7, Donmez2}). GRH equations are solved around the
non-rotating and rotating black holes in curved space-time which is described by Kerr
line element,

\begin{eqnarray}
    ds^2 = -(\alpha^{2}-\beta_{i}\beta^{i})dt^2  + 2\beta_{i}dx^{i}dt + \gamma_{ij}dx^{i}dx^{j},
\label{GRH2}
\end{eqnarray}

\noindent where $\alpha$ and $\beta_{i}$ are the lapse function and shift vectors, respectively, which are

\begin{eqnarray}
  \alpha = \left(1 + \frac{4Ma^{2}}{A} - \frac{2M}{r} \right)^{1/2},
\label{GRH3}
\end{eqnarray}

\noindent and 

\begin{eqnarray}
  \beta_{r} = 0, \;\;\;\; \beta_{\theta} = 0, \;\;\;\; \beta_{\phi} =  -\frac{2Ma}{r},
\label{GRH4}
\end{eqnarray}

\noindent where $A  = r^{4} + r^{2}a^{2} + 2Mra^{2}$.
$a$ is dimensionless black hole
spin parameter and $M$ is the mass of the black hole.

We model the torus' dynamic defined on the equatorial plane in the limit of an
infinitesimal thin disk. In fact, the tickness of the disk  is finite and icreasess
  during the accretion of the matter
  towards the black hole. The effect of the disk tickness on the non-axisymmetric
  instability, spectra, and the black hole spin would be small \citep{Zhou1}.
It is assumed that the black hole is kicked during
one of the astrophysical events processes. As a result of this kick,
the oscillating black hole
transfers the angular momentum to  the accreted torus, and
the stable torus is perturbed. In the case of kicked black hole, the matter around
the black hole receives an additional velocity.
It is  handled by focusing on perturbing the  radial
velocity of the torus  at every time step at the center
of galaxies or AGNs. The radial perturbations of the initially stable relativistic torus on
the equatorial plane around the
black holes due to the kicked black hole at the center of galaxies are
studied by updating the radial velocity of the torus 
with the following expressions:

\begin{eqnarray}
  v^r_{new} = v^r_{old} - \frac{\chi}{r^2}sin(\phi) \;\;\;\;\;\; when \;\;0
  \leq \phi \leq \pi \nonumber \\ 
  v^r_{new} = v^r_{old} + \frac{\chi}{r^2}sin(\phi) \;\;\;\;\;\; when \;\;
  \pi \leq \phi \leq 2\pi
\label{perturb velocity}
\end{eqnarray}

\noindent
where  $v^r_{new}$ and $ v^r_{old}$ are the velocities of torus' matter in the current and
previous time steps respectively, and
the amplitude of the oscillation is controlled by $\chi$. Hence, low and
high velocity kicks are
controlled by changing this parameter.

The non-self-gravitating relativistic torus was first studied analytically by \citet{Abramowicz1}.
A sharp cusp was found for marginally stable accretion disk. The disk was located on the equatorial
plane around the black hole. General relativistic torus with a perfect fluid equation of state was
numerically discussed in \citet{Zanotti1, Zanotti3, Zanotti2, Nagar1}.
They computed the
initial values of stable accreting tori around the non-rotating and the rotating black hole using $\Gamma=4/3$
to mimic a degenerate relativistic electron gas. In order to have a system in a
hydrodynamical equilibrium, the centrifugal and gravitational forces are balanced
by the internal pressure in the torus. In this paper, we study the non-stable black hole
torus system by perturbing the initially stable and non-self-gravitating relativistic
tori which were given in \citet{Zanotti1, Zanotti2, Donmez2}. Initial relativistic tori used
in our numerical simulations are
almost stable during the evolution without a perturbation as seen in Figs.9 and 10 in \citet{Donmez2}.
Initial physical parameters  of the stable torus and kicked parameters used throughout this paper are
summarized in Table \ref{table:Initial Models1}.

The initial setups of tori around the non-rotating and the rotating
black holes are thin, non-self-gravitating, and non-axisymmetric and defined
on the fixed space-time metric using the Kerr coordinate.
The disk-to-black hole mass ratio chosen as $\frac{M_t}{M} =0.1$
  in our initial models makes the torus-black hole system stable with respect to the
  runaway instability. Therefore these models only respond to perturbations with a small amplitude
  which creates instabilities on the torus \citep{Daigne1, Zanotti2}.
  Although the mass ratios  for the real accreting systems are significantly lower such as $X-$ray binaries and
  AGNs \citep{Putten2}, the instabilities created after the applied perturbation could show the same behavior due
  to the following reasons. First, the self gravity of the torus is neglected. Second, the fixed background metric
  will not be affected due to the falling matter into a black hole. Finally, our torus is located very close to the
  black hole horizon. The gravitational time scales would be
  much larger than the other time scales in our numerical domain. 
The $2D$ perturbation of the relativistic
torus on the equatorial plane is modeled in a polar coordinate. 
The inner and outer boundaries  of the computational domain
are extended from $r_{in}=2.8M$ to $r_{out}=100M$ around the non-rotating black hole.  It
goes from $r_{in}=1.7M$ to $r_{out}=100M$ for the rotating black hole.
The uniform grid is used in the all models along the radial and the angular
directions with $N_r=1024$ $\times$ $N_{\phi}=512$ cells.
The outflow boundary conditions are imposed at inner and outer boundaries of
computational domain.
The time evolution analysis of the
stable torus on a fixed space-time metric without any perturbation shows
that the structure of initial torus never changes  ( does not develop a instability) during
the simulation using the Cowling simulation
\citep{Cowling}. The neglecting
  the perturbation of the gravitational field is known as Cowling approximation. The
  Cowling approximation of the general relativistic system can be satisfied by setting
  the perturbation to be zero in the metric \citep{Shijn}. 
The cowling approximation guarantees having a
unperturbed gravitational field. The negligible mass accretion rates freeze the growth of instability
oscillation mode in the Cowling approximation \citep{Hawley1}.

\begin{table*}
\footnotesize
\caption{The physical parameters of initially stable accretion torus, black holes, and
  perturbation parameter used in our numerical simulations.
 \label{table:Initial Models1}}
\begin{center}
\vspace*{-2ex}
  \begin{tabular}{cccccccccc}
    \hline 
    \hline 
        &      &             &                  &    &    &   &  &  &   \\
    Model  & $\frac{a}{M}$  & $\frac{M_t}{M}$ & $\rho_c(geo)$ & $\ell_{0}$  & $r_{cusp}(M)$
    & $r_{lso}-r_c(M)$ & $t_{orb}(M)$ & $r_{out}(M)$ &$\chi (geo/M^2)$ \\
\hline
$K09A$   & $0.9$  & $0.1$ & $7.981x10^{-2}$ & $2.60$ & $1.78$  & $1.78-3.40$ & $39.4$ & $19.25$ & 0.01 \\
$K09B$   & $0.9$  & $0.1$ & $7.981x10^{-2}$ & $2.60$ & $1.78$  & $1.78-3.40$ & $39.4$ & $19.25$ & 0.001 \\
$K09C$   & $0.9$  & $0.1$ & $7.981x10^{-2}$ & $2.60$ & $1.78$  & $1.78-3.40$ & $39.4$ & $19.25$ & 0.0001 \\
$K00A$   & $0.0$  & $0.1$ & $1.140x10^{-4}$ & $3.80$ & $4.57$  & $4.57-8.35$ & $151.6$& $15.889$ & 0.01 \\
$K00B$   & $0.0$  & $0.1$ & $1.140x10^{-4}$ & $3.80$ & $4.57$  & $4.57-8.35$ & $151.6$& $15.889$ & 0.001 \\
$K00C$   & $0.0$  & $0.1$ & $1.140x10^{-4}$ & $3.80$ & $4.57$  & $4.57-8.35$ & $151.6$& $15.889$ & 0.0001 \\

\hline 
\hline 
  \end{tabular}
\end{center}
\end{table*}

   Main parameters of the initial stable torus around the black hole
  and its perturbation are given in Table \ref{table:Initial Models1}, including 
    the model name,  black hole spin parameter $a$,
    torus-to-hole mass ratio $M_t/M$, the maximum density of the torus $\rho_c(geo)$,
    constant specific angular momentum $\ell_{0}$,
    cusp location $r_{cusp}(M)$. 
    $r_{c}$ and $r_{lso}(M)$ represent the location of the maximum value of density and
    the inner radii of the last stable orbit respectively, and
    $t_{orb}$ is the orbital period at  $r = r_{c}$.  And $\chi (geo)$ is a
    free parameter to control the perturbation on the torus radial velocity.
    The adiabatic index $\Gamma =4/3$,  the polytropic constant $K(geo)= 4.969$ x $10^{-2}$
    is used in
    all simulations. We have used some negligible values for the initial density and pressure with
    a zero radial and angular velocities outside the torus.
    Mainly, we have constructed two different initial stable tori. First one is the torus
    around the rotating black hole and the second one is around the non-rotating black hole.
    The constant polytropic index  and specific angular momentum are used in all our models.

    The characterization and analysis of the instability are  done
    by using the Fourier mode analysis explained in \citet{Donmez2}. Mainly, we calculate the
    power mode by computing the real and imaginary parts in Fourier transfer in order to analyze the
    azimuthal mode $m$.


\section{Numerical Results}
\label{Numerical Results}

In this work, we have concentrated on a black hole-torus system which was perturbed by 
a kicked black hole, believed to occur at the center of AGN, and
have conducted the dynamical calculations of the kicked torus.
We neglect the self-gravity of the torus material and magnetic
field along with using fixed space-time metric around the rotating and the non-rotating
black holes.
The immediate change in the kinetic energy of the torus' matter due to the kick would
adjust the new equilibrium configuration. The change would also cause the transferring
angular momentum of the torus which liberates the gas falling toward or away from the
black hole.
The redistribution of the angular momentum of the accreted torus due to
the oscillating black hole allows the newly formed structures of the disk to evolve.
During the evolution, the mass would be transferred, which can lead to forming an
spiral arms and a  quasi-periodically oscillation  of the disk. 

\subsection{Rotating Black Hole}
\label{Rotating Black Hole}

In order to study the effect of the kicked black hole onto the resulting torus dynamical
evolution, we perform the number of numerical simulations around the rotating black hole
for the different values of the perturbation parameter $\chi$. The torus' matter could
escape from the gravity of the black hole or remain the bound depending on the
direction of the kick and distance.

\begin{figure*}
  \center
 \includegraphics[width=0.31\textwidth]{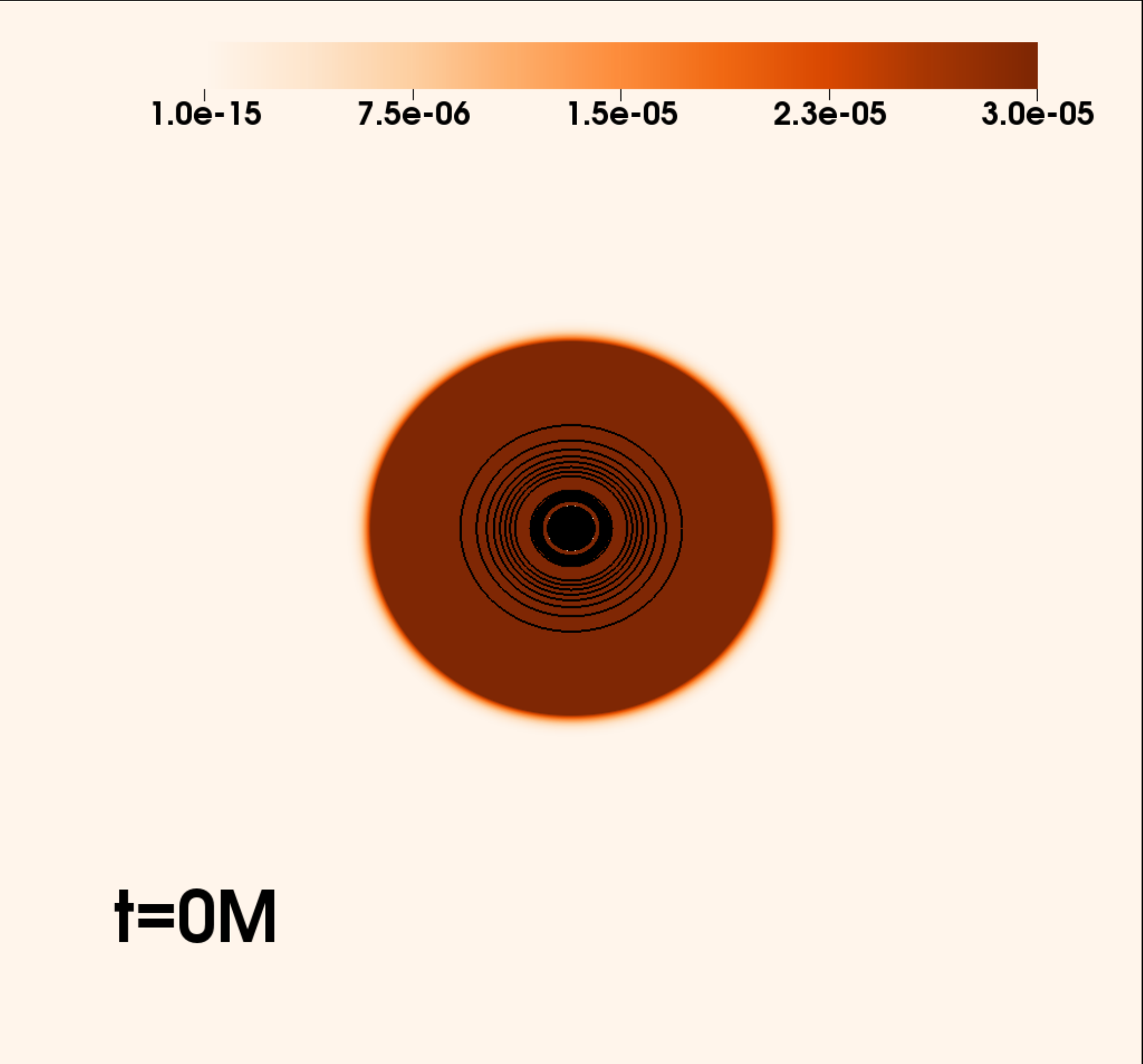}
 \includegraphics[width=0.31\textwidth]{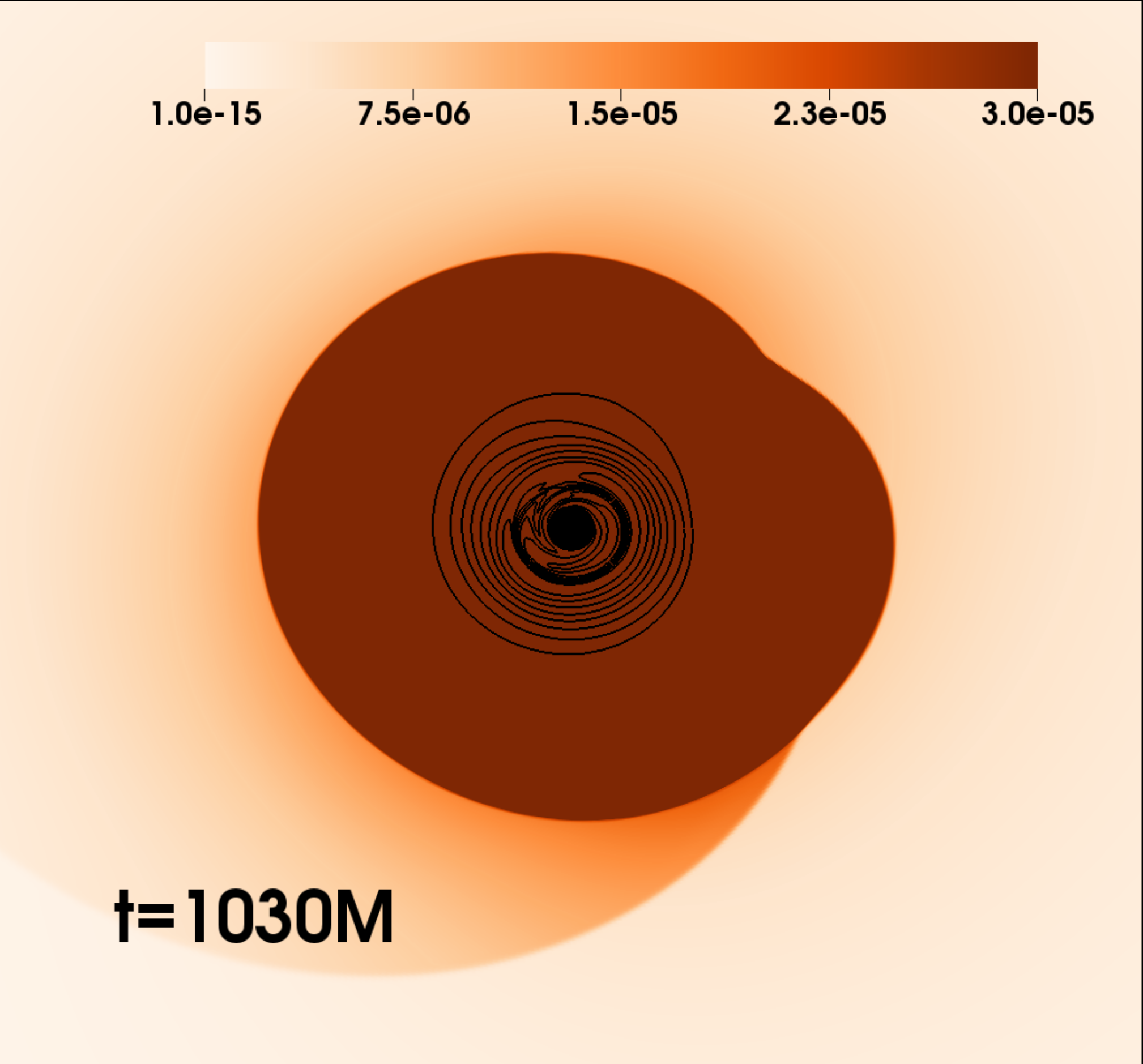}
 \includegraphics[width=0.31\textwidth]{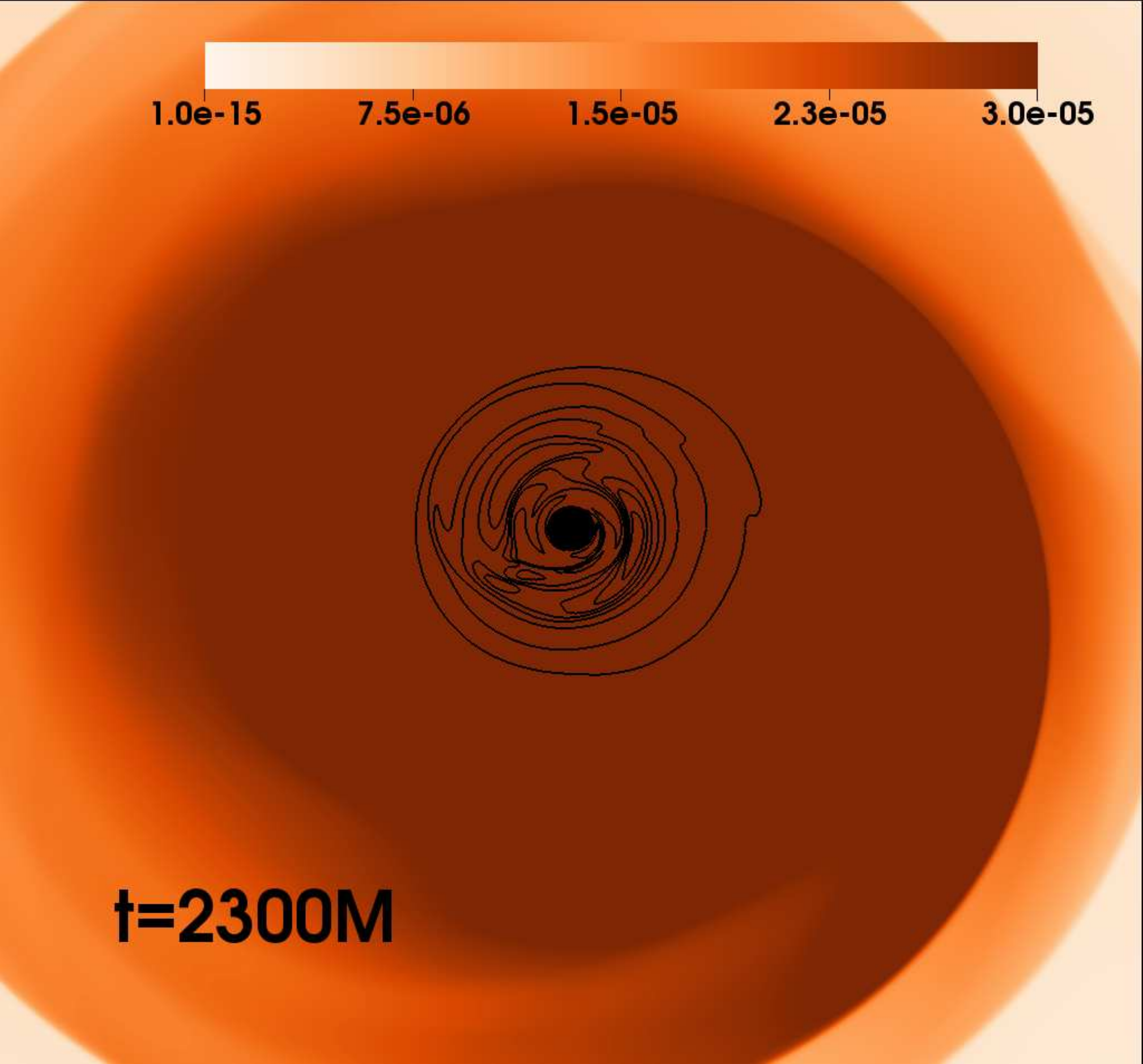}
 \includegraphics[width=0.31\textwidth]{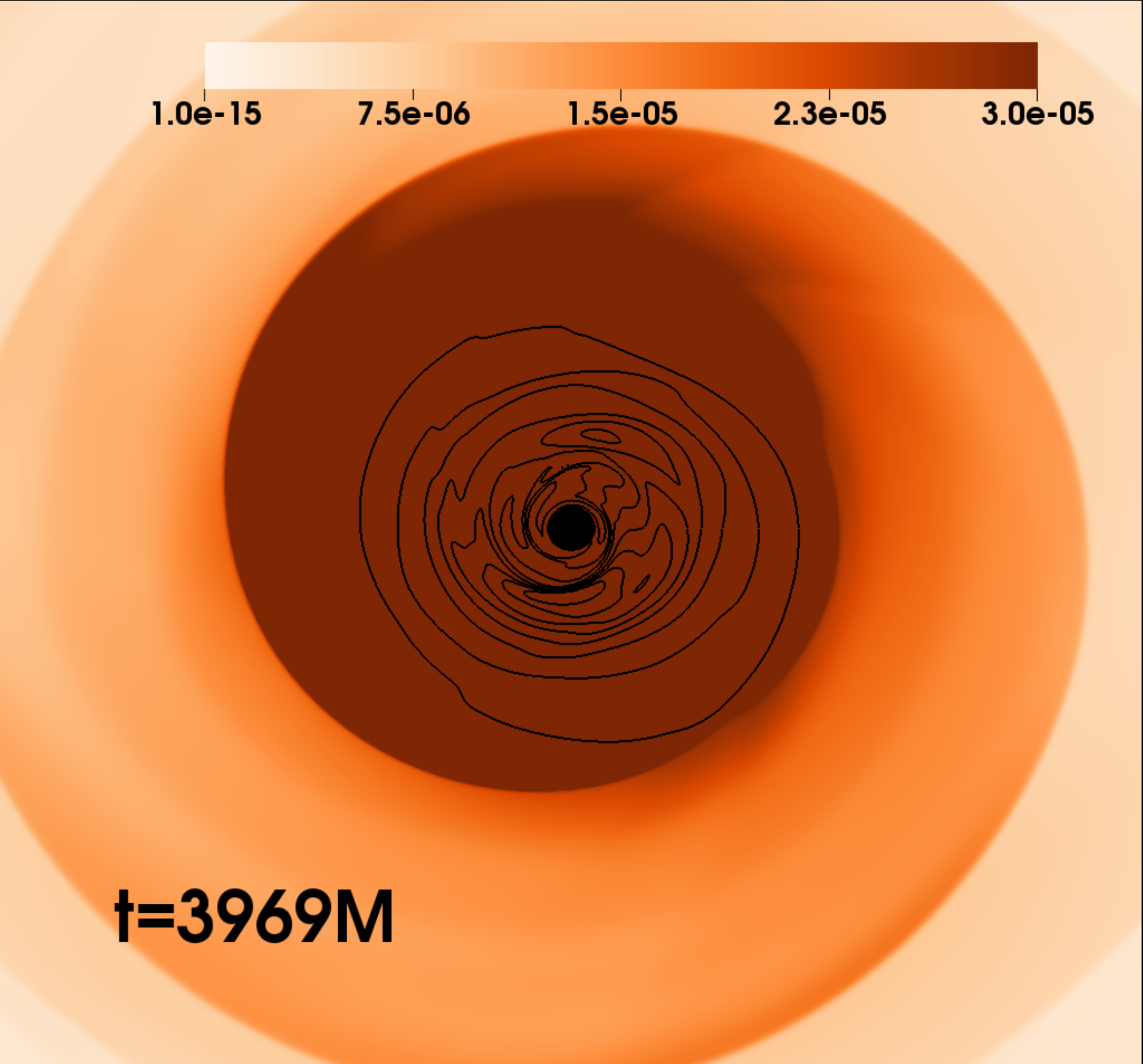}
 \includegraphics[width=0.31\textwidth]{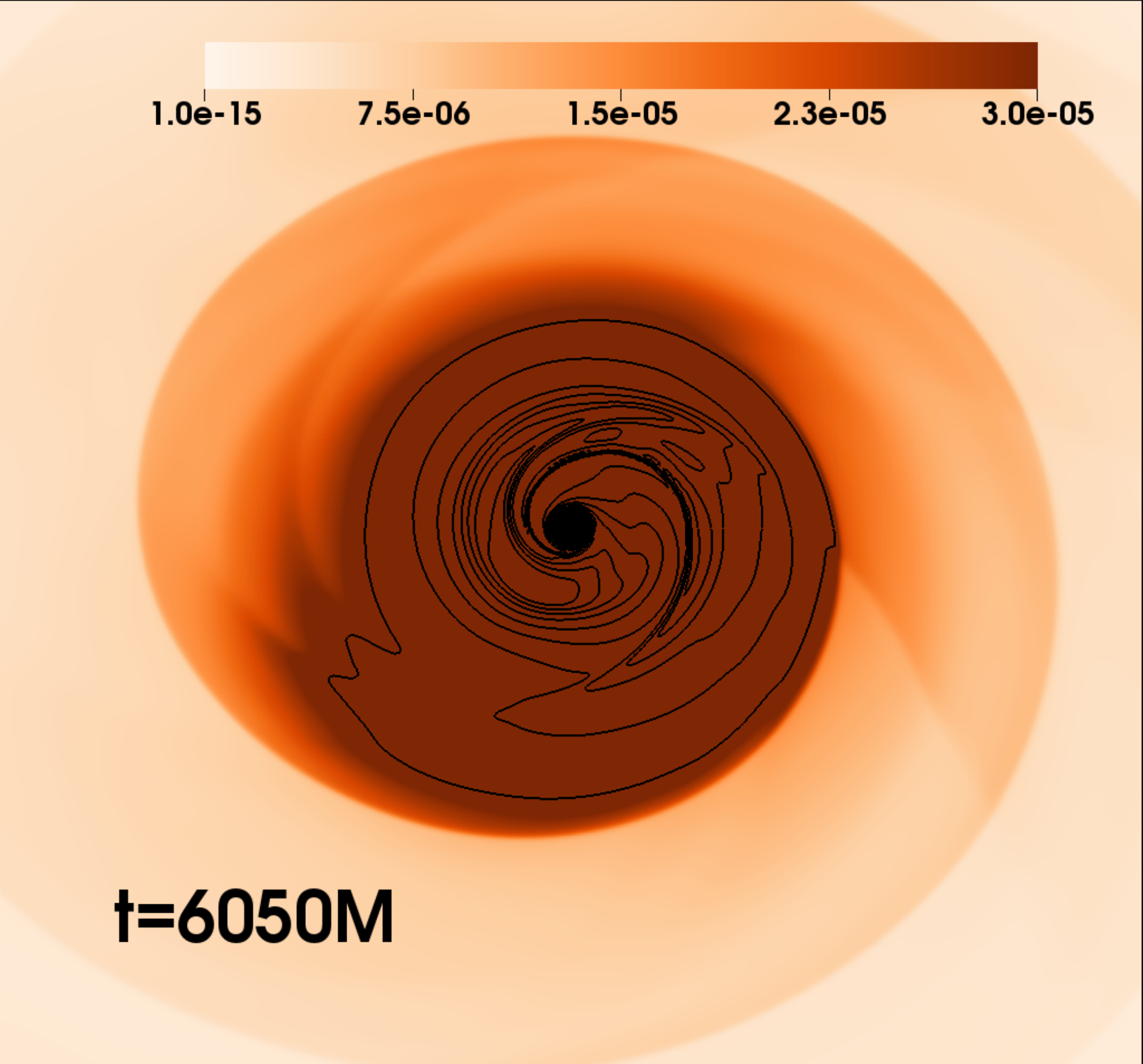}
 \includegraphics[width=0.31\textwidth]{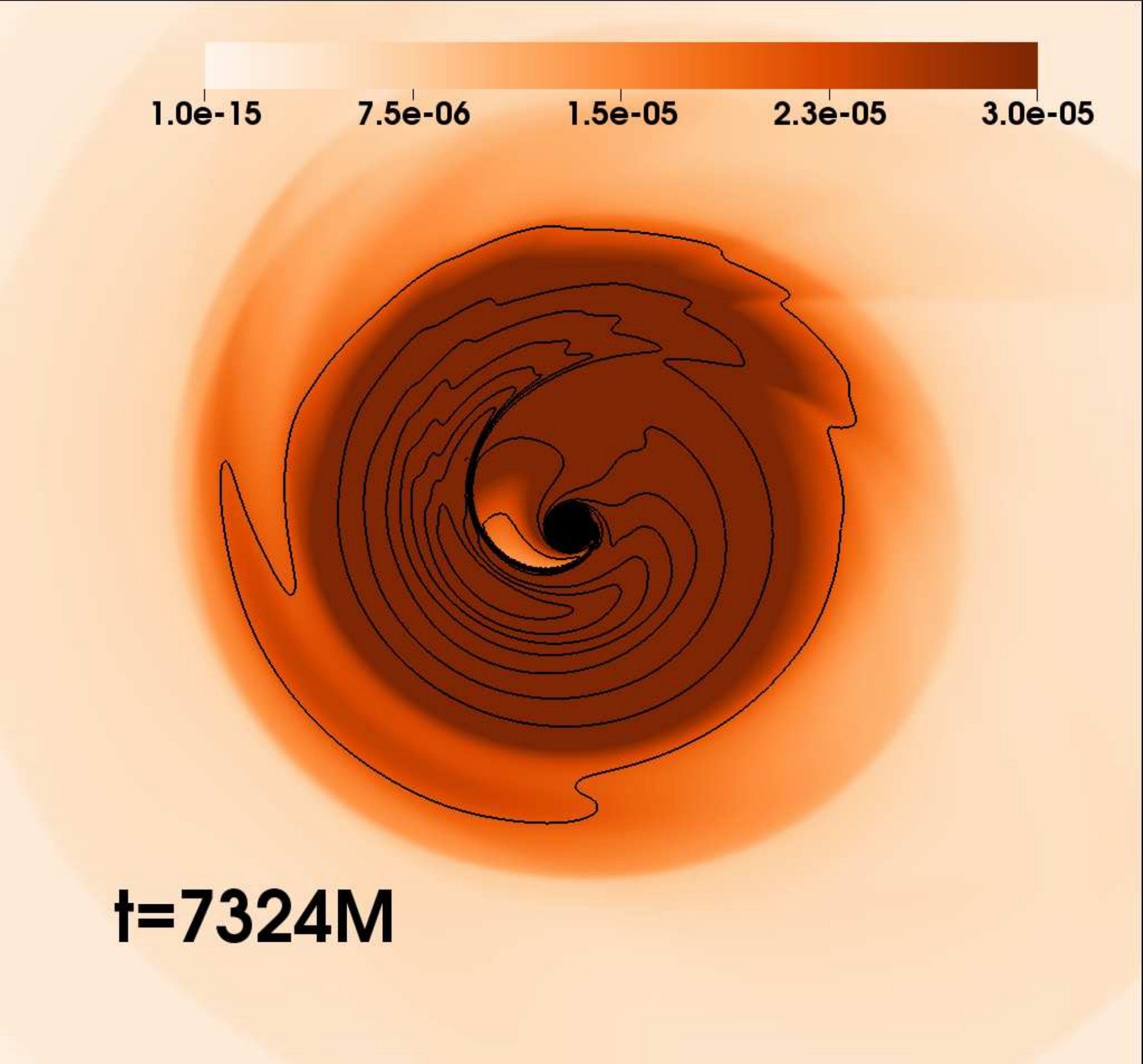}
 \includegraphics[width=0.31\textwidth]{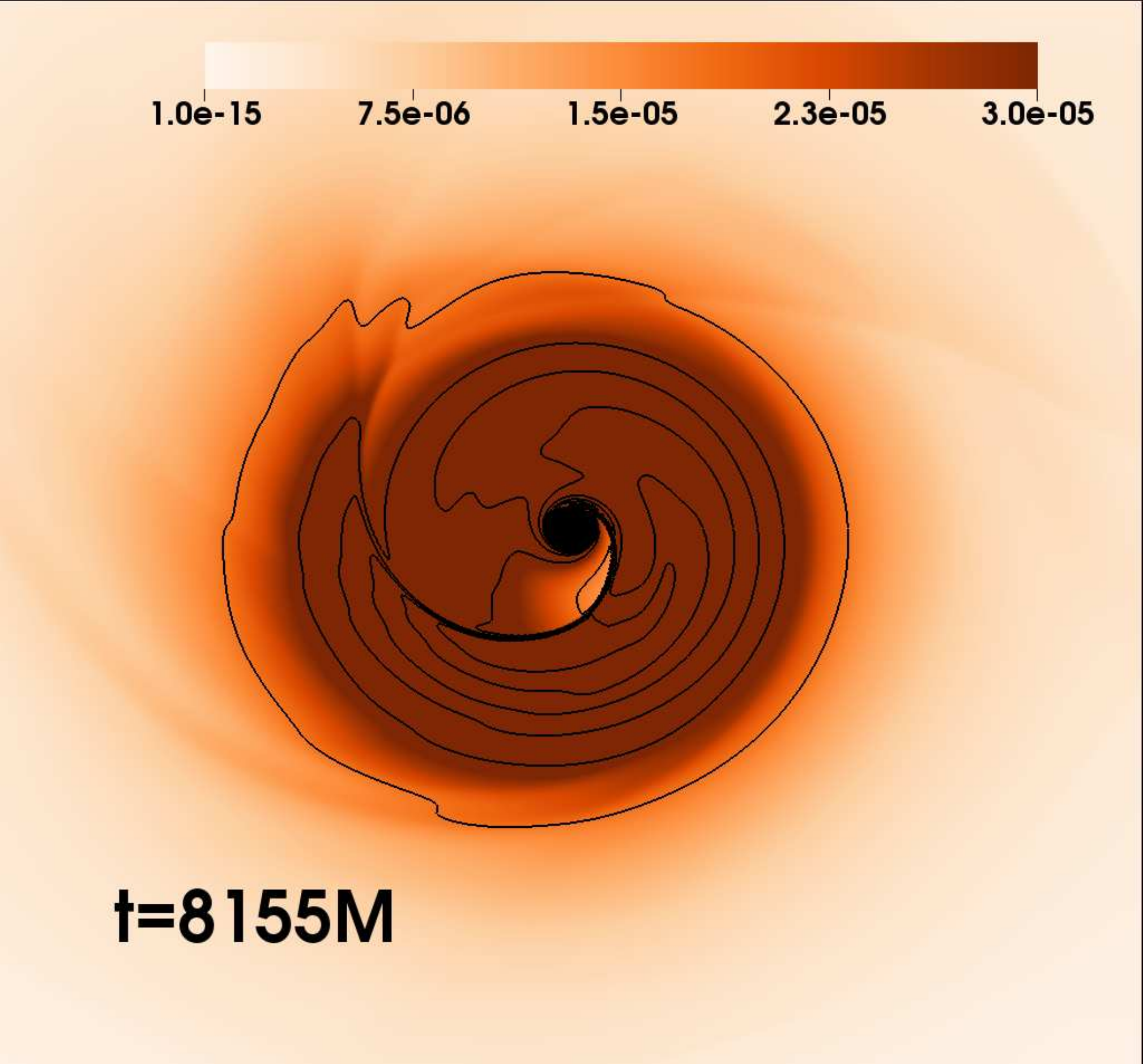}
 \includegraphics[width=0.31\textwidth]{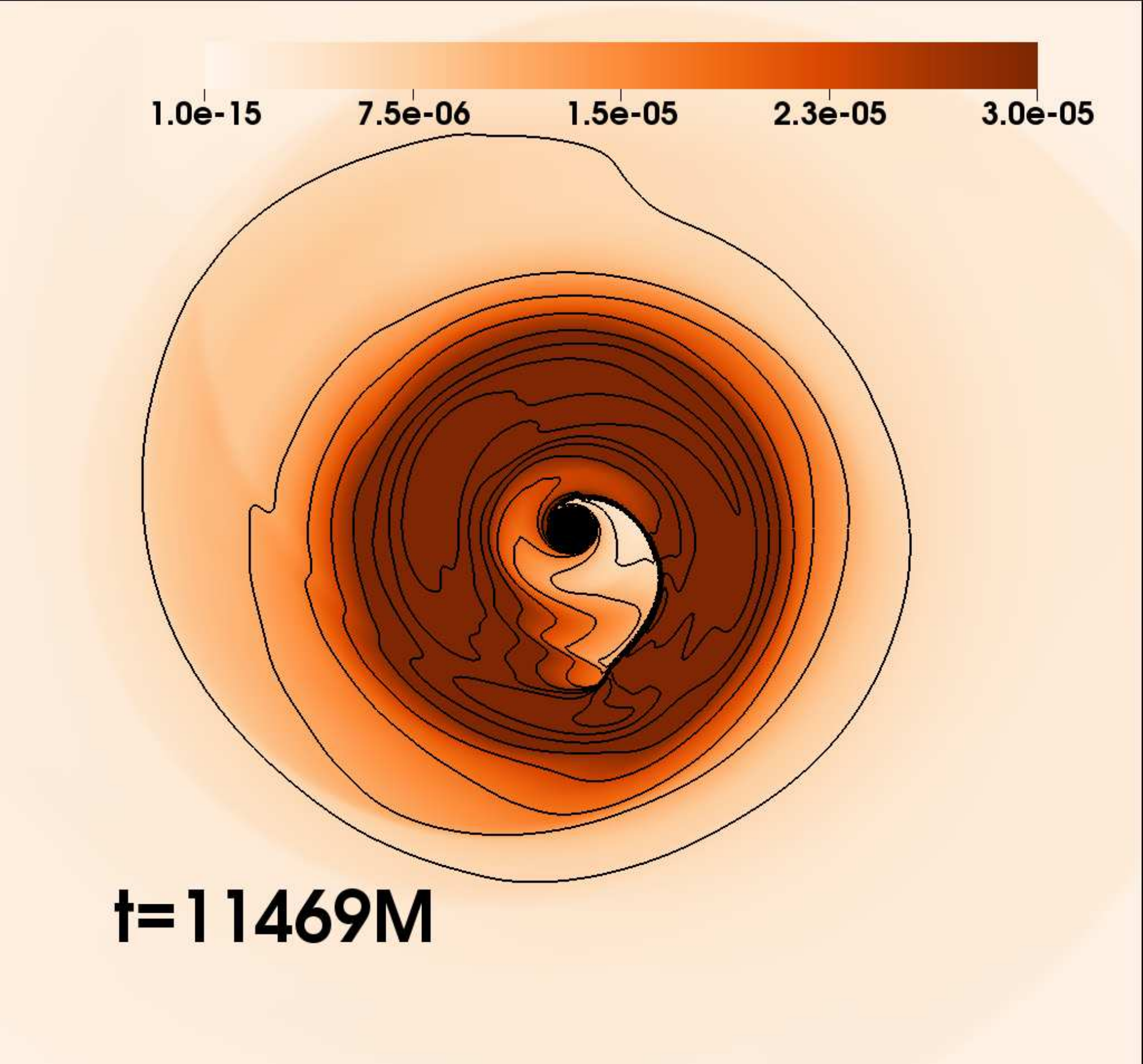}
 \includegraphics[width=0.31\textwidth]{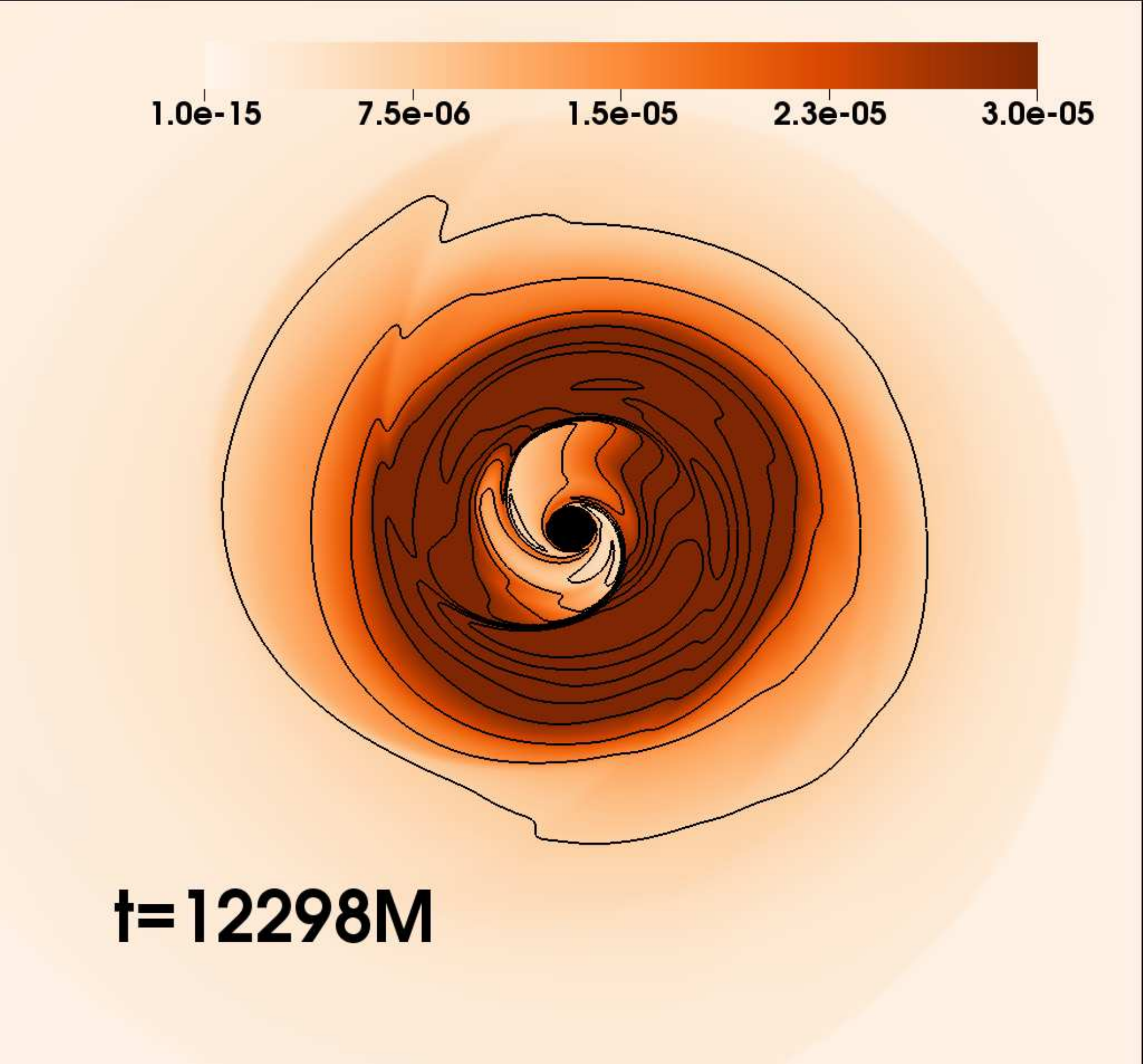}
 \includegraphics[width=0.31\textwidth]{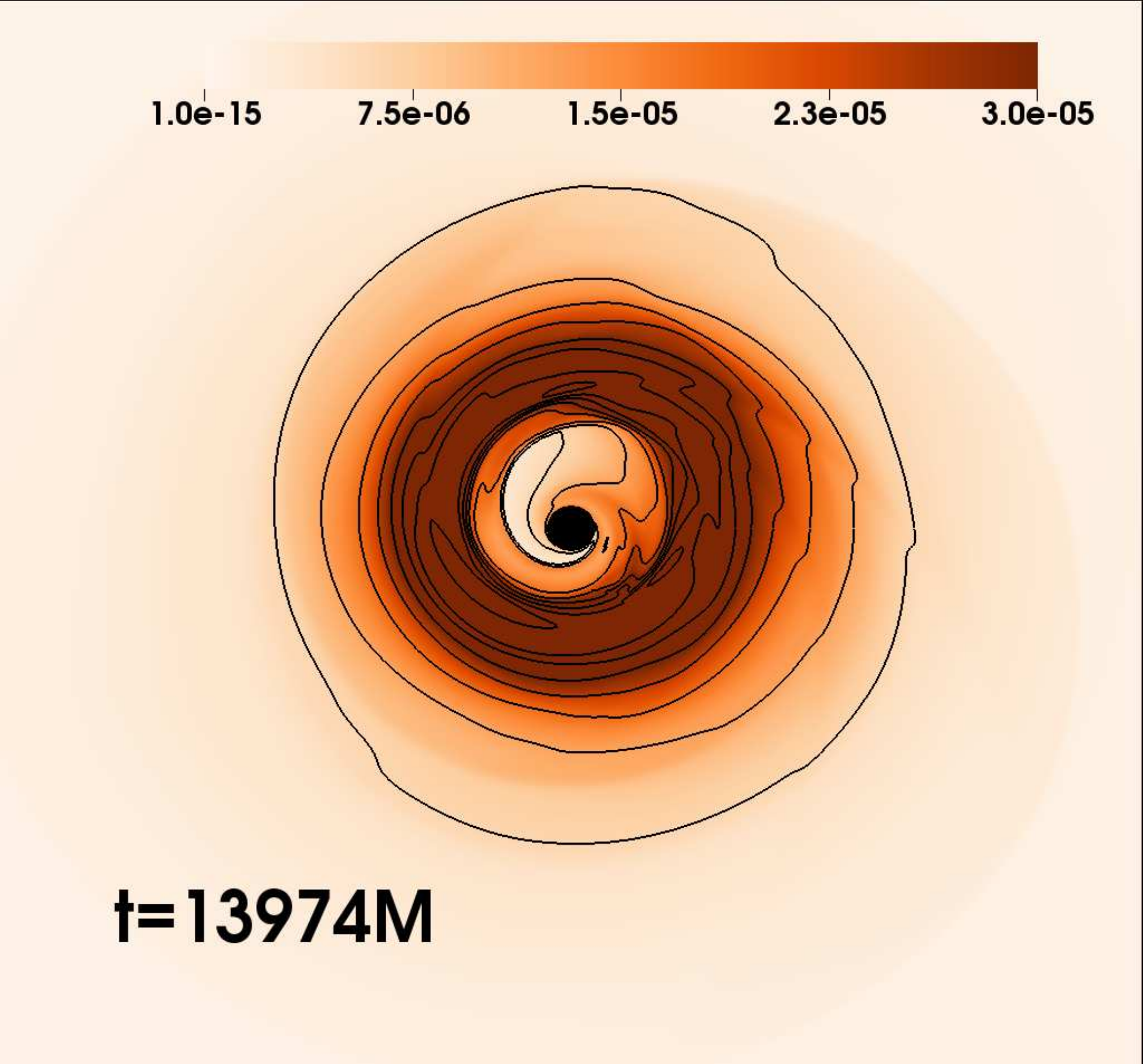}
 \includegraphics[width=0.31\textwidth]{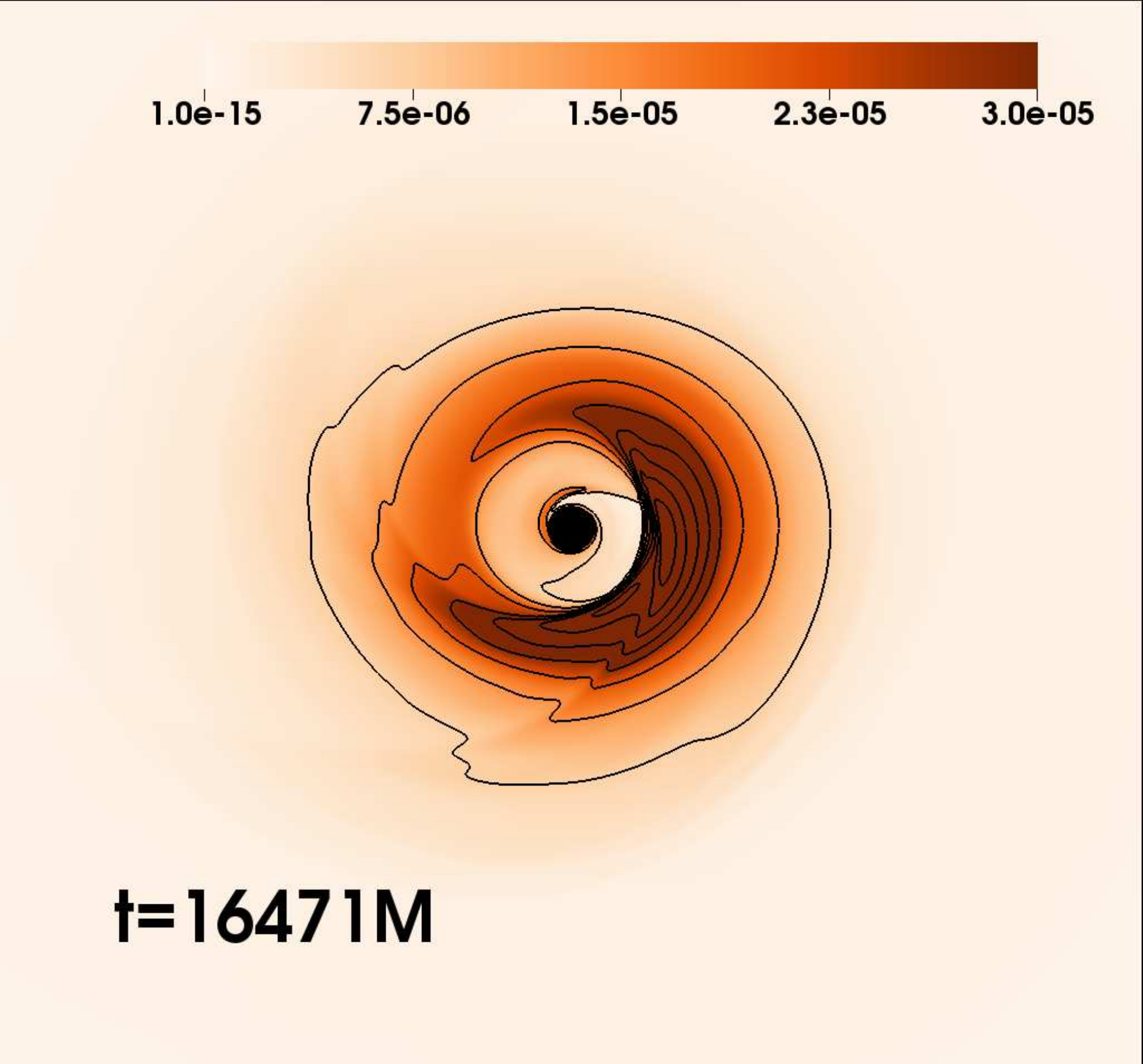}
 \includegraphics[width=0.31\textwidth]{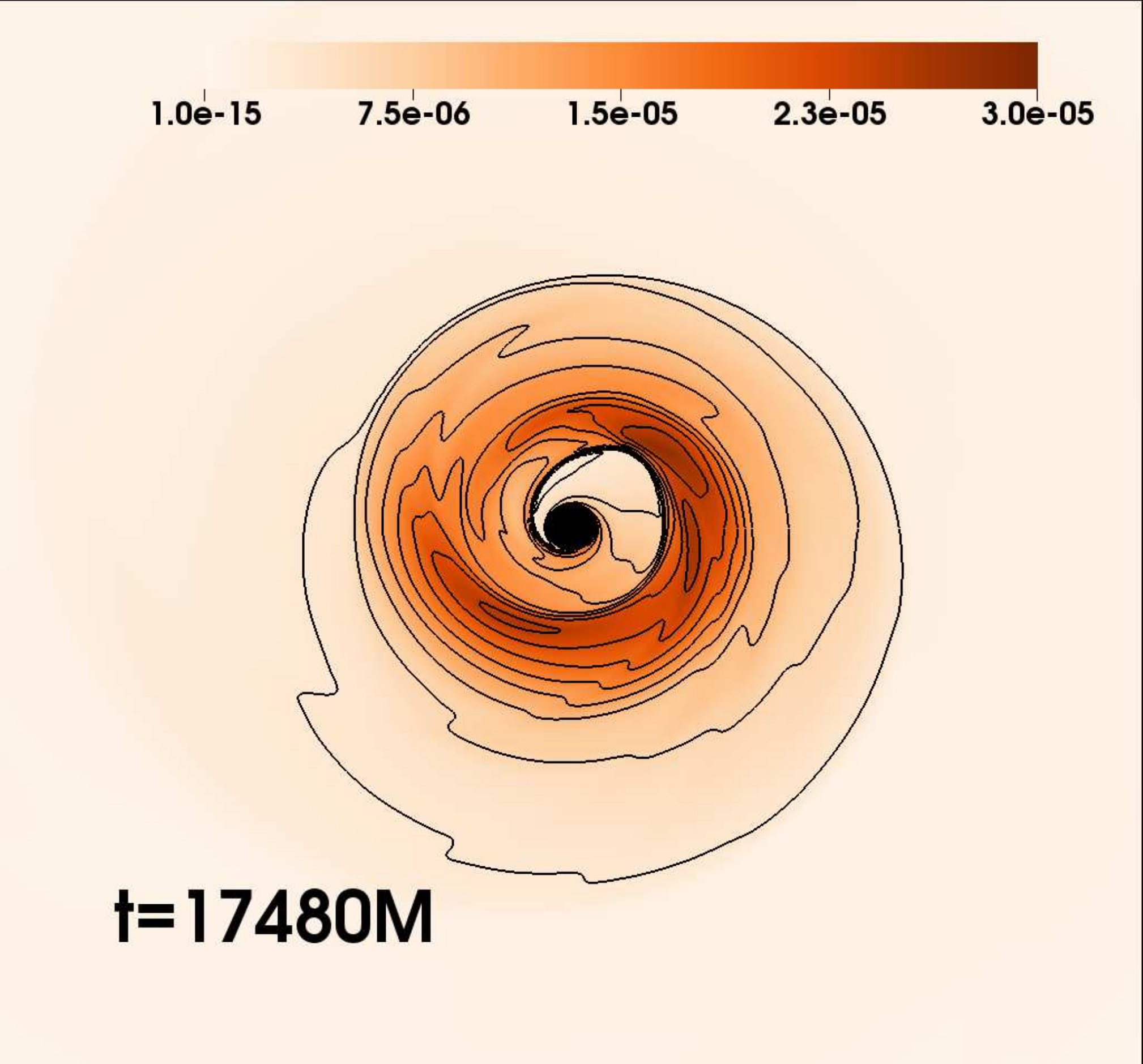} 
 
 \caption{The rest-mass density of the torus on
   equatorial plane with linearly spaced
   isocountours around the rotating black hole for the model $K09C$.
   The spiral structure seen in snapshot is a
   mechanism to have a $m=1$ mode.
   {\bf The domain is $[X_{min},Y_{min}] \rightarrow [X_{max},Y_{max})] =
     [-40M,-40M)]\rightarrow [40M,40M]$.}}
\label{rotating_1}
\end{figure*}

Morphology and dynamical evaluation of perturbed torus around the
rotating black hole for model $K09C$ is given in Fig.\ref{rotating_1}. This figure
represents the evaluation of the logarithmic rest-mass density on the
equatorial plane at different snapshots. The last snapshot given in the figure
shows the distribution of matter after the matter has almost completed $446$
orbital periods around the black hole.  The orbital period of the models
for each model is given in Table \ref{table:Initial Models1}.
As it is seen from the snapshots,
perturbation due to the kicked black hole initially triggers the spiral arms around
the equilibrium. The instability   is extended all over the  torus
surface. The oscillating torus, seen in Fig.\ref{rotating1_4}, starts filling the region
and induces a small accretion process towards the black hole. This process does not create
a big shock wave to reduce the rest-mass density of the torus until the snapshot at $t=6070M$,
as seen in Fig.\ref{rotating_1}. Later, the oscillation amplitude of the torus is getting bigger
during the time evolution, and it causes the creation of the spiral shock wave. The spiral wave causes
the angular momentum transfer and the matter the falling towards to or away from the black
hole. Having a shock wave also causes kinetic energy to convert to thermal energy. The spiral shock waves
can be clearly seen in Fig.\ref{rotating_1} starting from the snapshot $t=7324M$.

The spiral density wave traveling with same angular velocity of matter on the torus creates co-rotation radius.
Inside this radius, matter travels slowly when it is compared with outside. Therefore,
negative angular momentum
appears  inside the wave which  causes the matter to fall
towards the black hole.

\begin{figure*}
  \center
  \includegraphics[angle=90,width=0.7\textwidth]{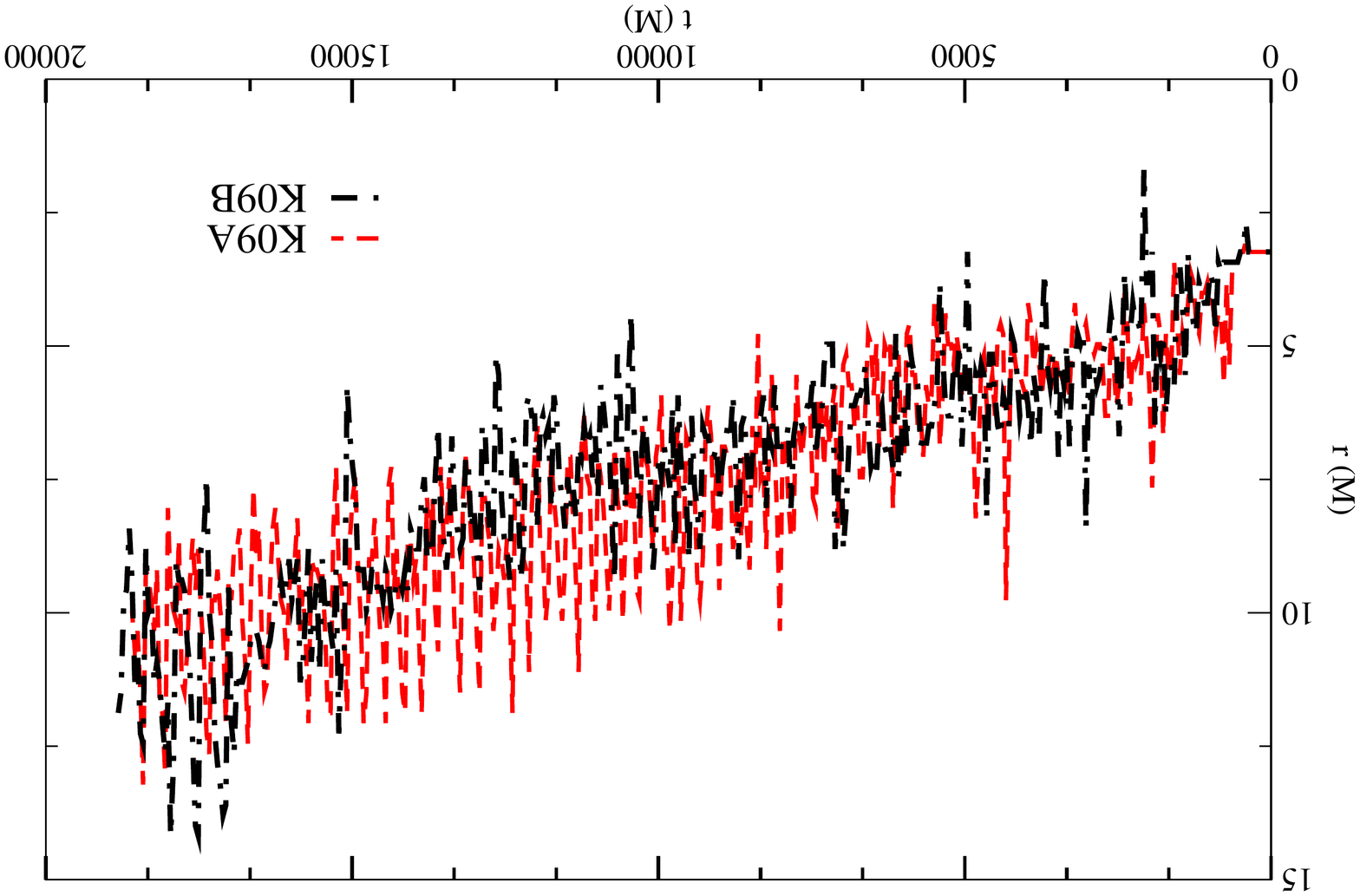}
\caption{The locations of the maximum rest-mass density for models
  $K09A$ and $K09B$.
  It gives the position
  of the maximum rest-mass density at the radial distance as a function of time.}
  \label{rotating_3}
\end{figure*}

\begin{figure}
 \center
\includegraphics[width=0.7\textwidth]{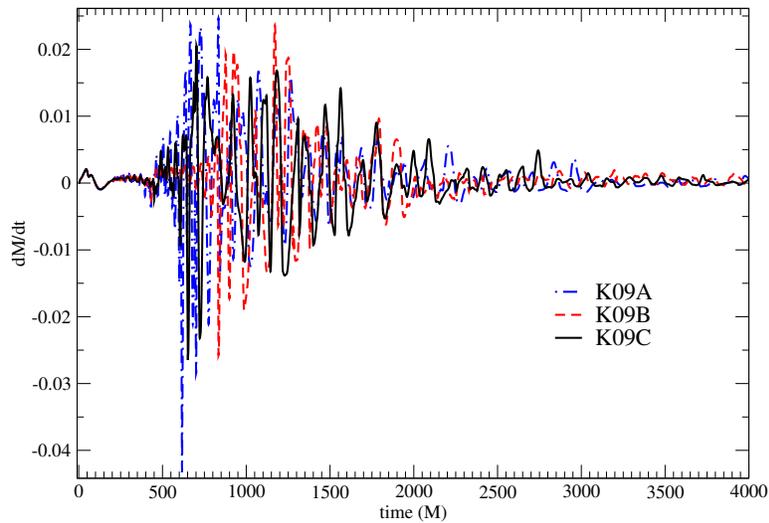}
\caption{The mass accretion rate at $r=3.8M$ for the models $K09A$, $K09B$, and $K09C$.}
\label{rotating1_4}
\end{figure}

The position of the maximum rest-mass density in the disk during the evolution is
shown in Fig.\ref{rotating_3} around the black hole for different models. As seen from the
figure, the effect of kicked black hole onto the disk  dynamic is almost the same for
different kick values which are represented with $\chi$ in Table \ref{table:Initial Models1}.
However, the positions of the maximum densities make nonlinear oscillation during the evolution.
The oscillation might be appeared due to the growth of the gravitational instability (spiral) mode.
On the other hand, the maximum rest-mass density of the stable torus  was initially at $3.4M$.
After the perturbation is applied, the position of the maximum rest-mass density is pushed out from
the black hole horizon with a significant distance  $r \sim 9M$ around $t \sim 10000M$.

Transferred angular momentum from the kicked black hole to the torus starts to
affect the torus dynamics after a few dynamical periods $(\sim 4t_{orb})$.
Initial perturbation causes instability on the dynamic of the torus and
the smaller the transferred angular momentum, the more instability develops
in later time of simulation seen in Fig.\ref{rotating_3} for the rotating black hole.
A higher black hole spin creates a larger centrifugal barrier  so that
it causes to be less accretions towards to the black hole. 

The mass accretion rates in arbitrary unit around the rotating black hole in case of different
perturbation velocities are shown in Fig.\ref{rotating1_4}. It is computed at inside the torus
at $r=3.8M$ close to the black hole horizon.
While the accretion rate $dM/dt > 0$ represents the perturbed torus' matter falling
  towards the black hole, the matter moves away from the black hole for $dM/dt < 0$.
  Fig.\ref{rotating1_4} has a clear indication of  the torus' matter oscillates due to black hole kick.
All the three cases in Fig.\ref{rotating1_4}
indicate that after significant oscillation phase like sinusoidal functions, the mass accretion rate
tends to a fairly constant value. It is found that the amplitude of the perturbation makes some delays
in case  of having an instability but these perturbations do not strongly responsive to
accretion rate in overall.

The torus with a constant specific angular momentum around the black hole is unstable in the 
non-axisymmetric global mode \citep{Chen1}.
The non-axisymmetric global mode is produced
on the torus due to kicked black hole. In order to the reveal the growth of this global mode
$m=1$, we compute the power mode amplitude  from the density of torus. Power mode is basically
computed by finding the Fourier decomposition of the azimuthal distribution of the tours density
during the time evolution
by using the following equation \cite{Donmez2}
\begin{eqnarray}
 P_m = \frac{1}{r_{out}-r_{in}} \int_{r_{in}}^{r_{out}}{ln ([Re(w_m(r))]^2+[Im(w_m(r))]^2) dr},
\label{Mod1}
\end{eqnarray} 

\noindent
where $r_{in}$ and $r_{out}$ are inner and outer boundaries of the computational domain,
respectively. The real and imaginary parts are
$Re(w_m(r)) = \int_{0}^{2 \pi}\rho (r,\phi) cos(m \phi) d\phi$ and \\
$Im(w_m(r)) = \int_{0}^{2 \pi}\rho (r,\phi) sin(m \phi) d\phi$.

\begin{figure}
 \center
\vspace{0.3cm}
\includegraphics[width=0.7\textwidth]{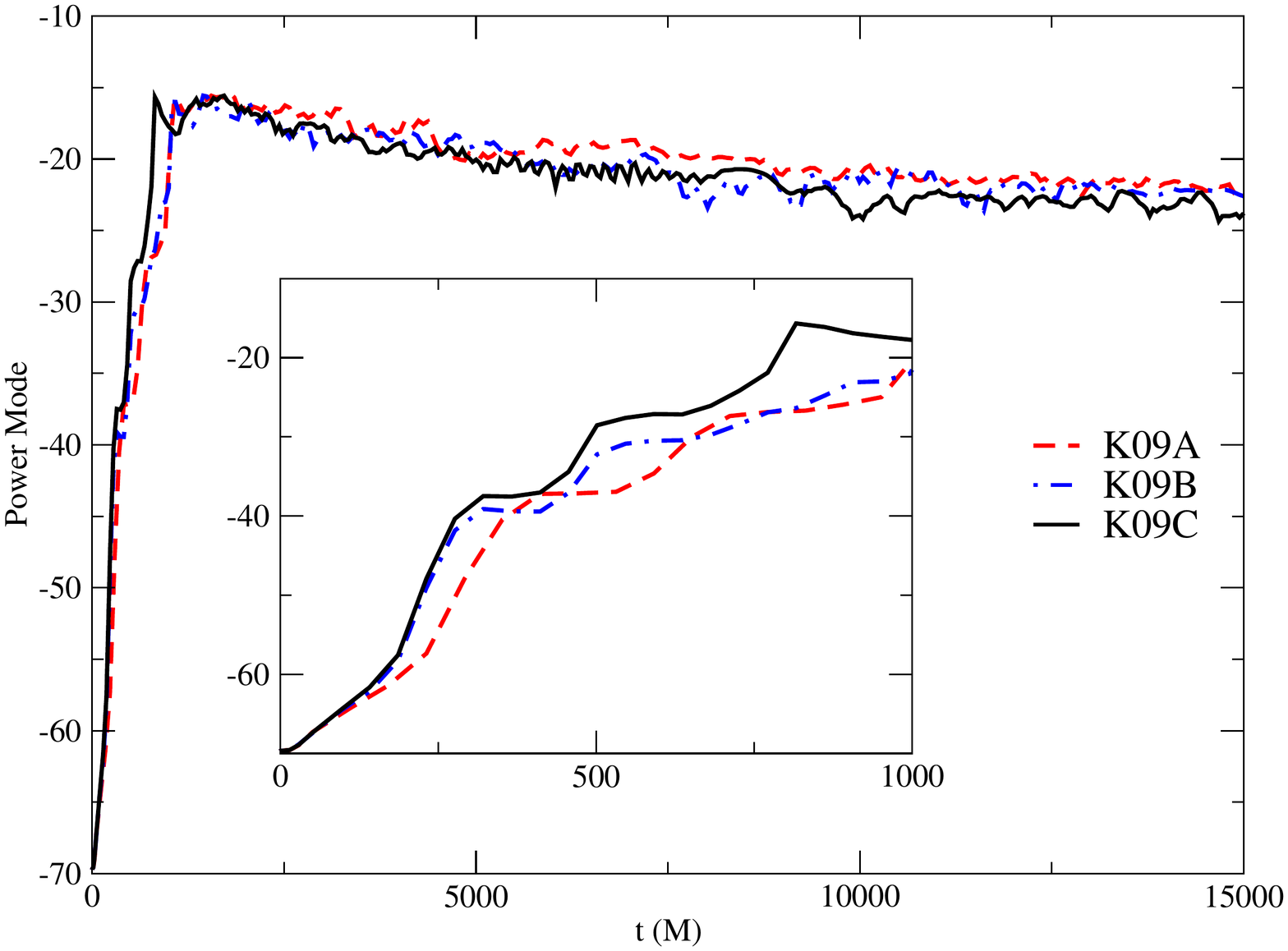}
\caption{Power mode evolution of $m=1$ non-axisymmetric mode for
  the models $K09A$, $K09B$, and $K09C$. It saturates around $t=1350M$ and the saturation
  amplitude is almost constant during the time evolution. The quasi periodic
  frequencies could be determined after the saturation point.  
\vspace{0.3cm}
\label{rotating1_5}}
\end{figure}

As seen in
Figs.\ref{rotating1_4} and \ref{rotating1_5}, the magnitude of
perturbation influences the process
of stabilization of the matter and saturation time because the angular momentum
of the initially stable torus faces a big influence due to the kicked black hole.
It is clearly found that the the density wave seen for $m=1$ in exponential growth mode
is developed around the rotating black hole. It is shown
in Fig.\ref{rotating1_5} that the saturation time  is slightly different for different perturbation
velocities. The higher the perturbation velocity, the more delayed it is to reach the saturation point.
However, the  amplitudes of mode $m=1$ for different models are almost the same. The growing in instability
can happen just by oscillating the black hole-torus system.

\begin{figure*}
  \center
 \includegraphics[width=0.31\textwidth]{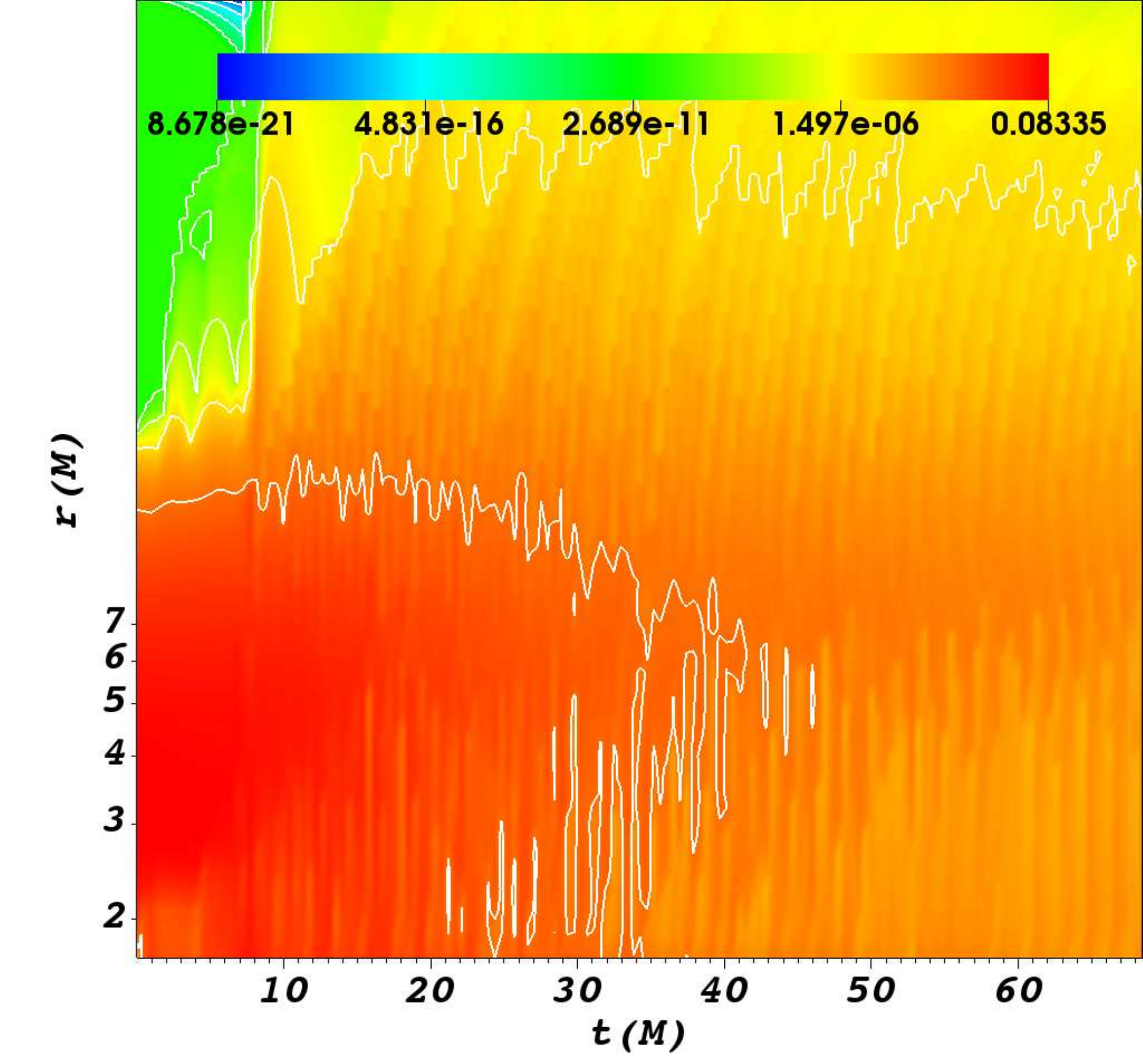}
 \includegraphics[width=0.31\textwidth]{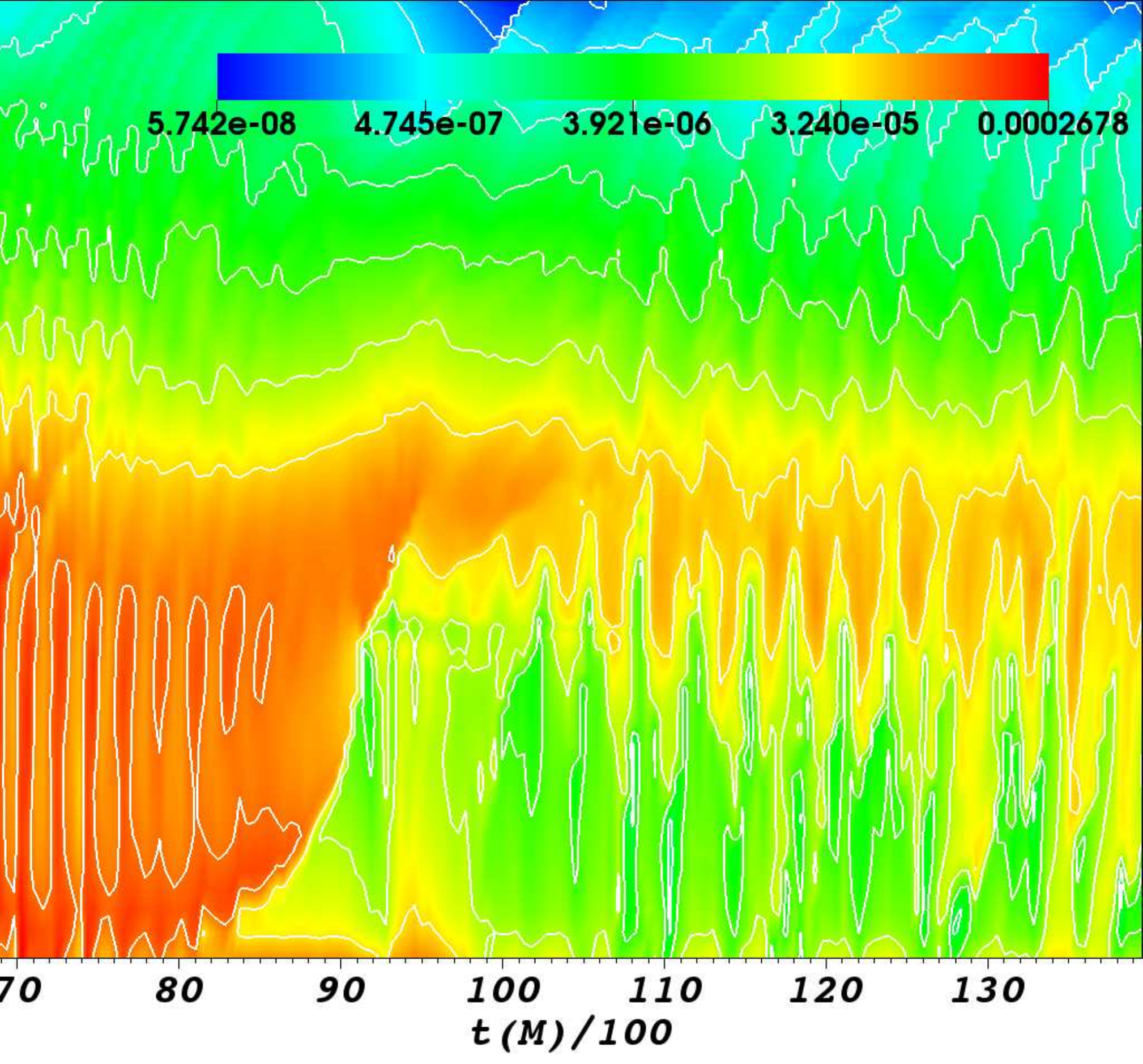}
 \includegraphics[width=0.31\textwidth]{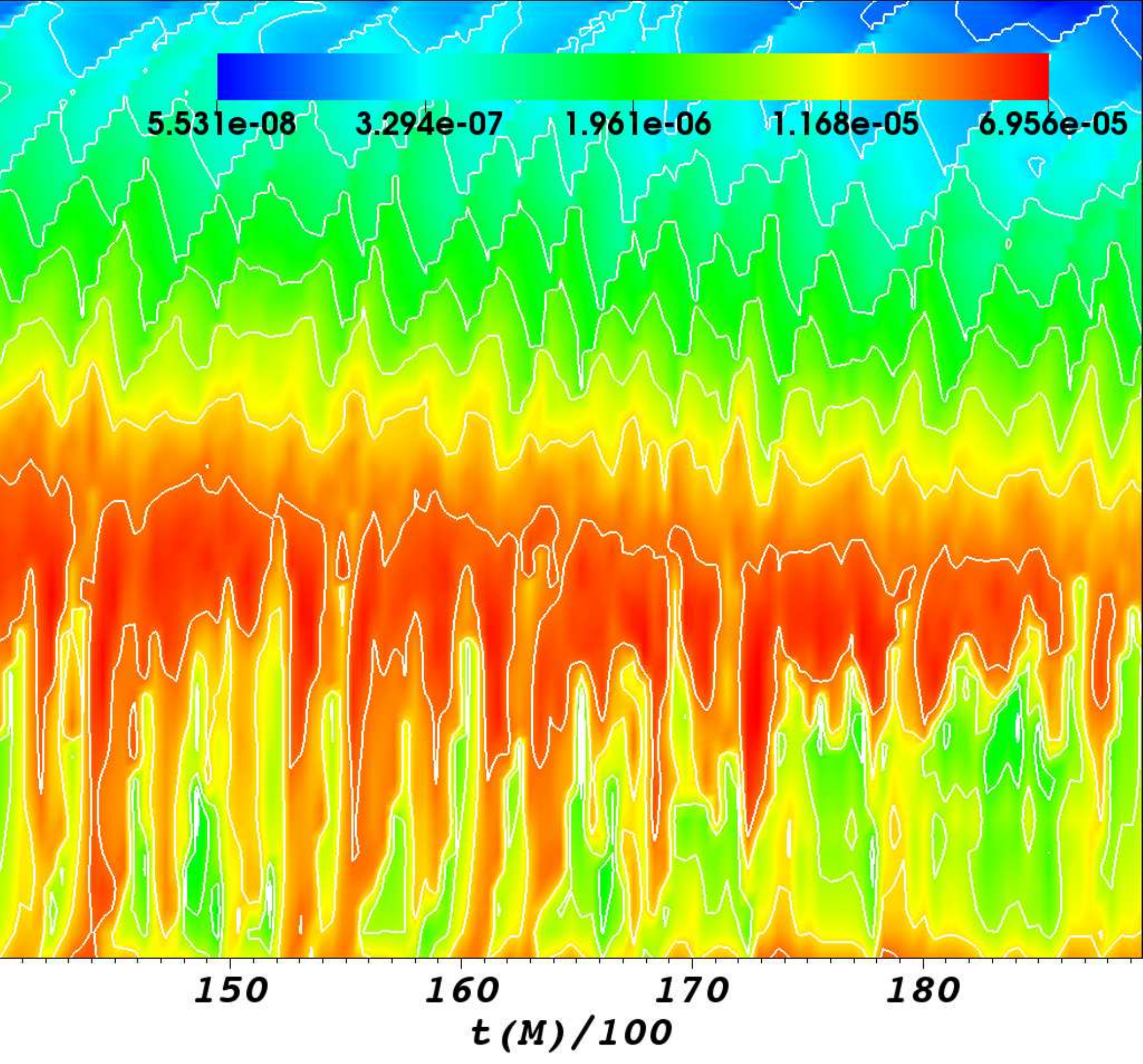} 
 \caption{The maximum value of the logarithmic density of the perturbed torus for model
   $K09C$ at fixed $\phi=0.0245$ is plotted as a
   function of time. The time axis is separated into three snapshots and each snapshot shows
  a different range of times within the simulation which enables us to see more details about the
   internal structure of the disk
   dynamics, especially in later times
   of the simulation. The kicked black hole creates an instability
   in the inner region of torus and it causes
   the torus' matter
   either falling into the black hole or going away from the black hole.}
\label{rotating_6}
\end{figure*}

\begin{figure*}
  \center
 \includegraphics[width=0.31\textwidth]{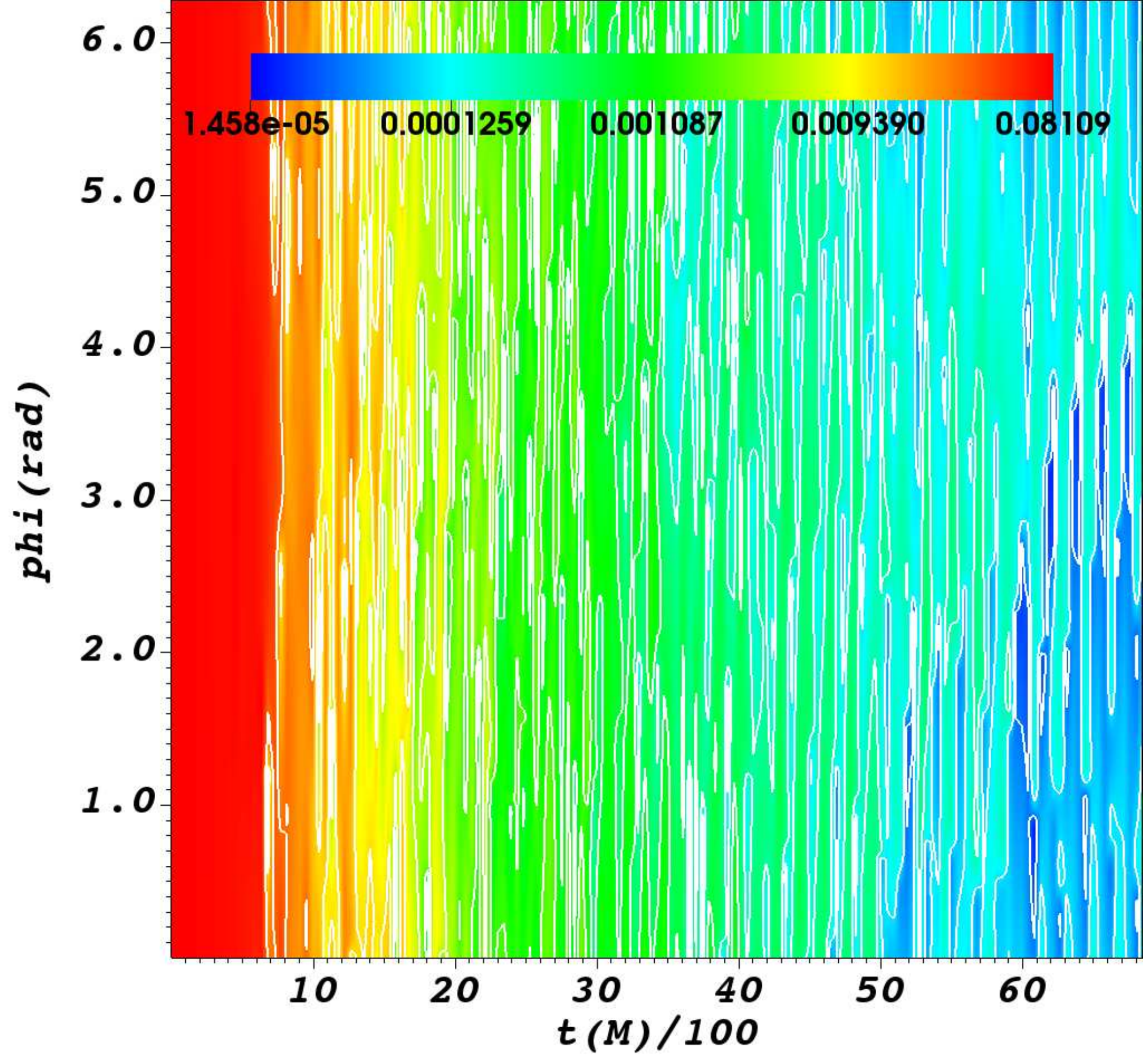}
 \includegraphics[width=0.31\textwidth]{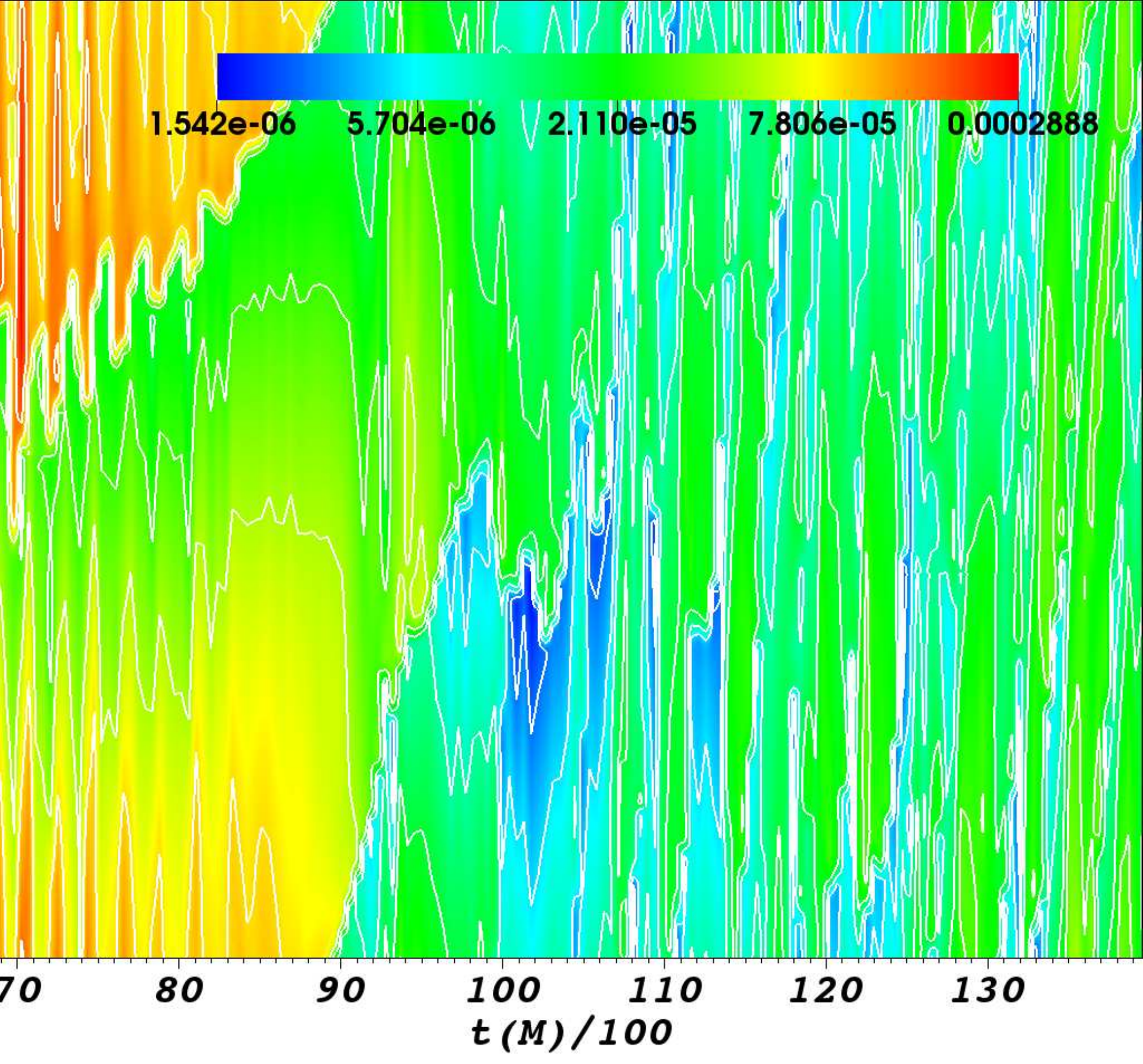}
 \includegraphics[width=0.31\textwidth]{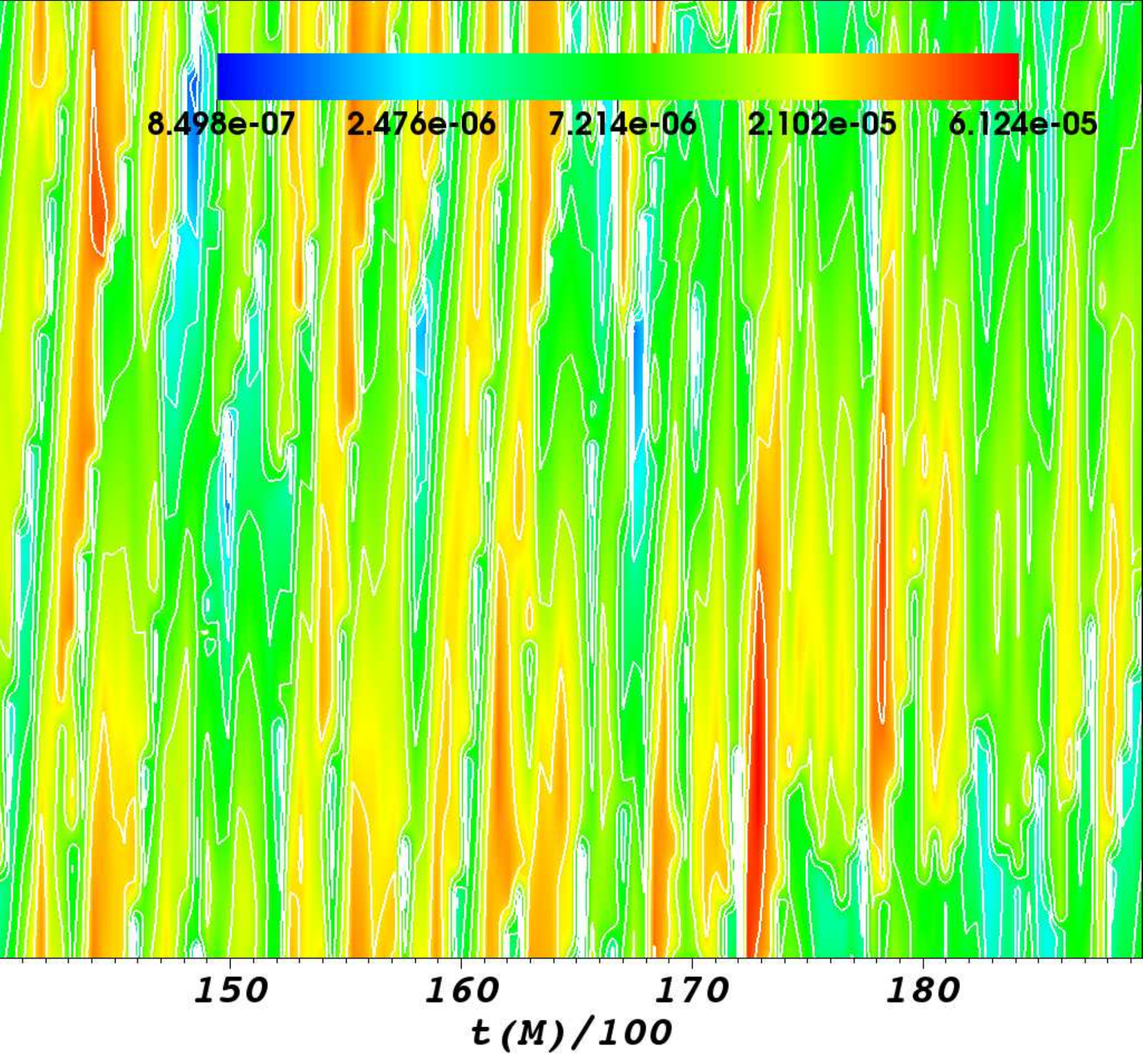}  
 \caption{Same as Fig.\ref{rotating_6} but it is for fixed $r=3.42M$.
   phi $(\phi)$} represents the angular direction.
\label{rotating_7}
\end{figure*}

The space time diagrams for  $(\phi - t)$  and $(r - t)$  of the
torus logarithmic rest-mass density during the time evolution
around the rotating black hole to see how and where the spiral arms
are created are shown in Figs.\ref{rotating_6} and \ref{rotating_7}, respectively.
In order see more detailed dynamics of the accreted torus, whole evolutions of the numerical
simulations are separated into three pieced. In the early stages of the simulation for model $K09C$,
the maximum value of density of
torus would not be effected too much from
the perturbation due to the  kicked black hole, seen in Figs.\ref{rotating_6} and \ref{rotating_7}.
However, the spiral arms are created during the evolution and this would lead to  the collimation of
the radiation into beams.
These types of phenomena  could be used to explain why the super Eddington luminosity is observed close to
the black hole horizon. The spiral shocks created the hot corona in the
inner region of the accretion disk  due to
frictional and compressional heating. The accretion due to spiral shock varies highly
with time and this type of accretion is responsible for the irregular behavior and the quasi-periodic oscillation
observed in black hole-torus systems in galaxies and AGN.
As it is seen in the middle panel of Fig.\ref{rotating_6}, the oscillation is  excited around the maximum
    of the epicyclic frequency which is $10M$ in the rotating black hole.
Additionally, high energetic
interactions inside the spiral arms would scatter emission from the region close to the black hole. 

%

\begin{figure*}
  \center
\includegraphics[width=0.7\textwidth]{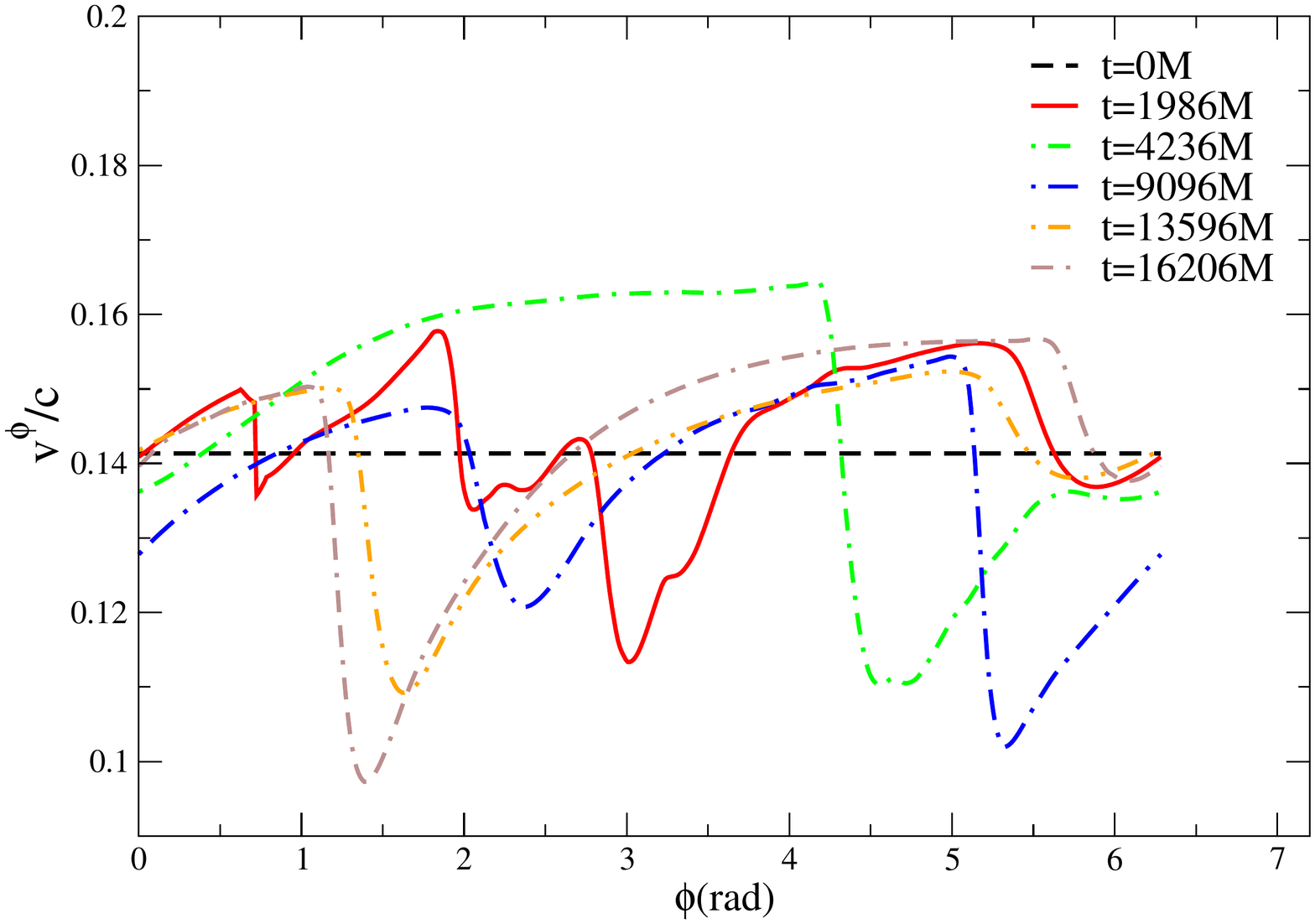}
\includegraphics[width=0.7\textwidth]{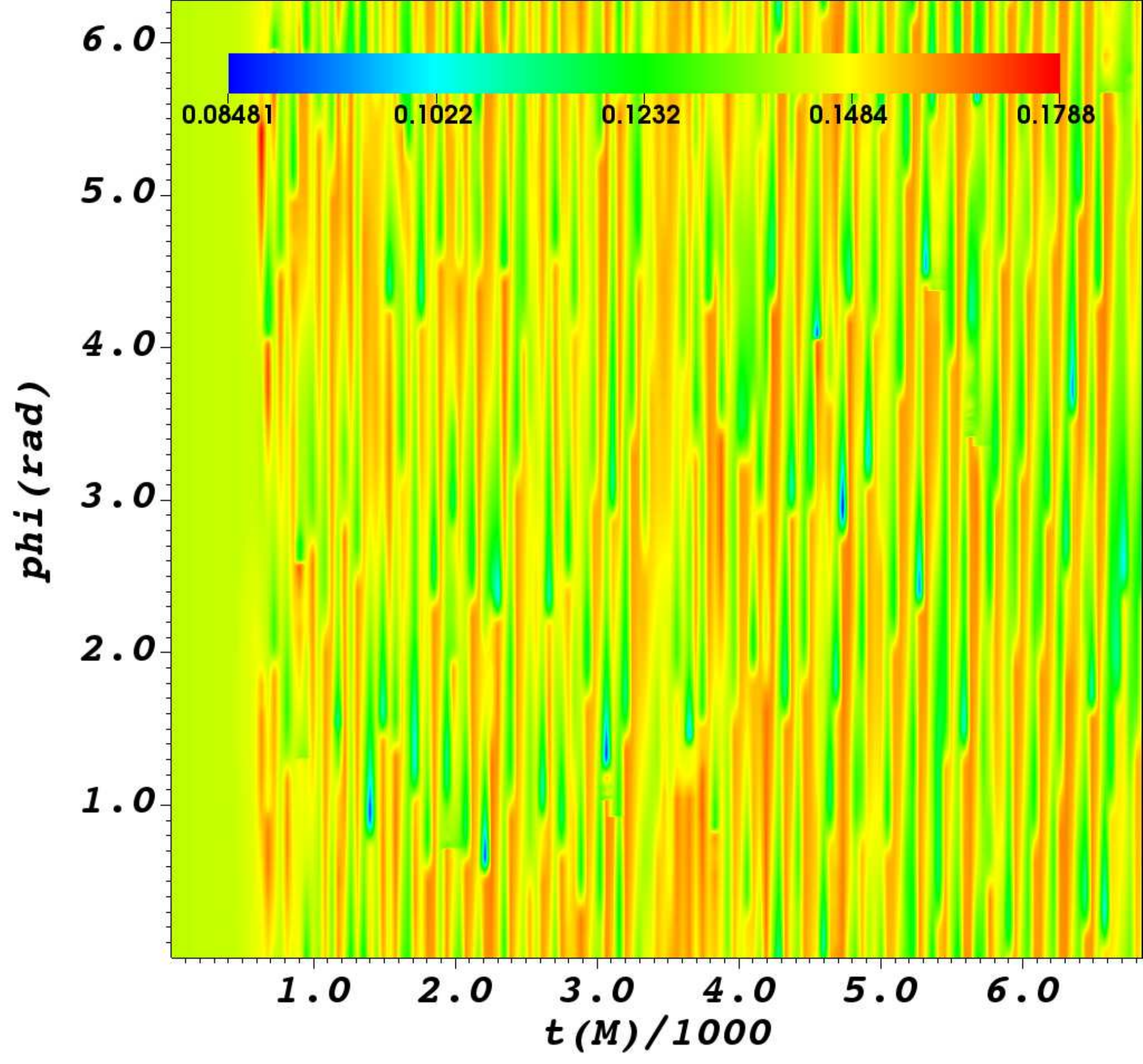} 
 \caption{Top panel: Angular profile of the three-velocity along the angular direction
   phi $(\phi)$
   with different times at $r=3.42M$ for model $K09A$. While the black dashed curve
   shows the angular velocity for the initial torus, the others represent the evolution
   of the velocity during the time. Bottom panel: The logarithmic angular velocity  is
   represented by a surface whose colors display the variation of the spiral shock
 along the angular direction.}
\label{rotating_8}
\end{figure*}

Initial torus with a constant specific angular momentum is rotating
around the black hole and perturbed by the black hole itself. These types of
perturbations would cause a redistribution of the specific angular momentum of the
torus. In order to mimic the  time evolution of the redistribution of the
specific angular momentum and clarify the effects of spiral shocks, we plot
the variation of angular velocity in Fig.\ref{rotating_8}. As it is seen in the top panel
of the figure, the instability grows early time of simulation $t\sim 300M$, the angular velocity is
redistributed and it causes rotating a strong spiral shock which can be seen after $t=1986M$.
From  our  numerical simulations  it  is  obvious  that  the  influence  of perturbation
on angular velocity creates a strong spiral shock which is excited in the disk
and propagate through the stable circular orbit.
The stronger the spiral shock would lead
the matter the accreting  toward the black hole and might be very important to explain the spectral
properties of the black hole candidates. 
Bottom panel of Fig.\ref{rotating_8}
has a clear evidence about the creation of a strong shock at certain times. The angular velocity
is varying between the smallest and the highest values of the color bar at $t\sim 1500M$,
$t\sim 2200M$, $t\sim 3100M$, $t\sim 4700M$, etc.

%


\subsection{Non-rotating Black Hole}
\label{Non-rotating Black Hole}

\begin{figure*}
  \center
 \includegraphics[width=0.31\textwidth]{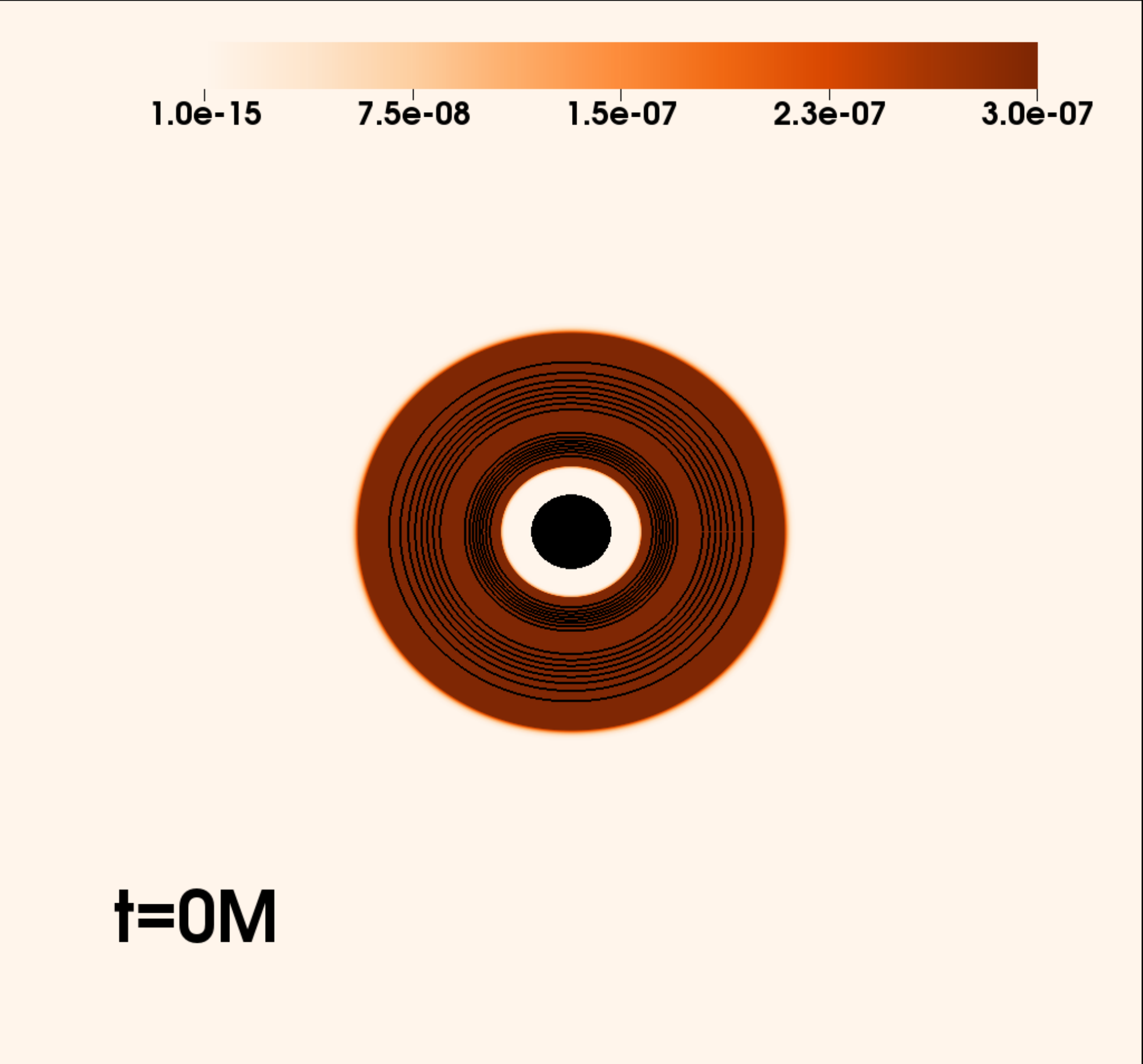}
 \includegraphics[width=0.31\textwidth]{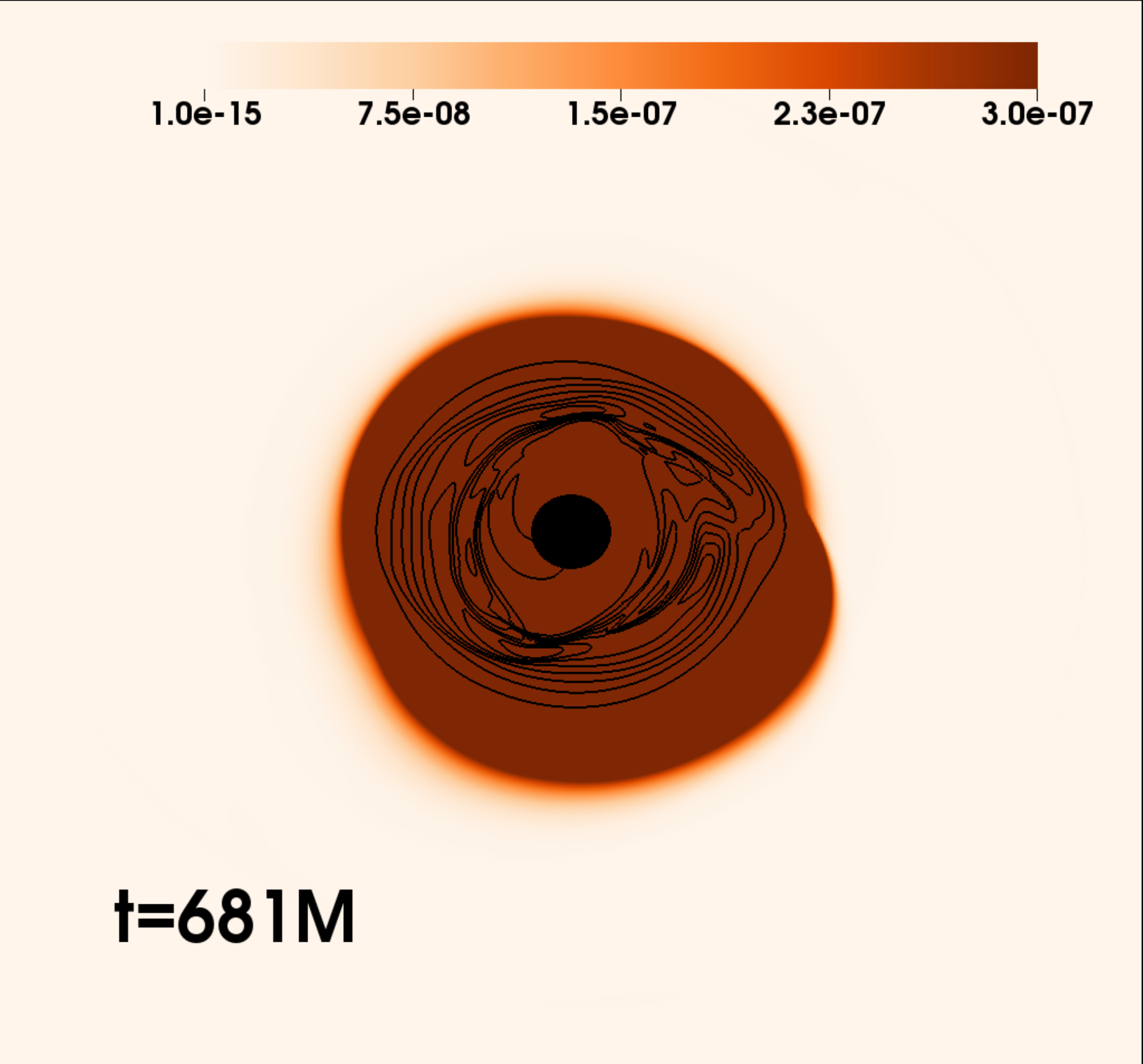}
 \includegraphics[width=0.31\textwidth]{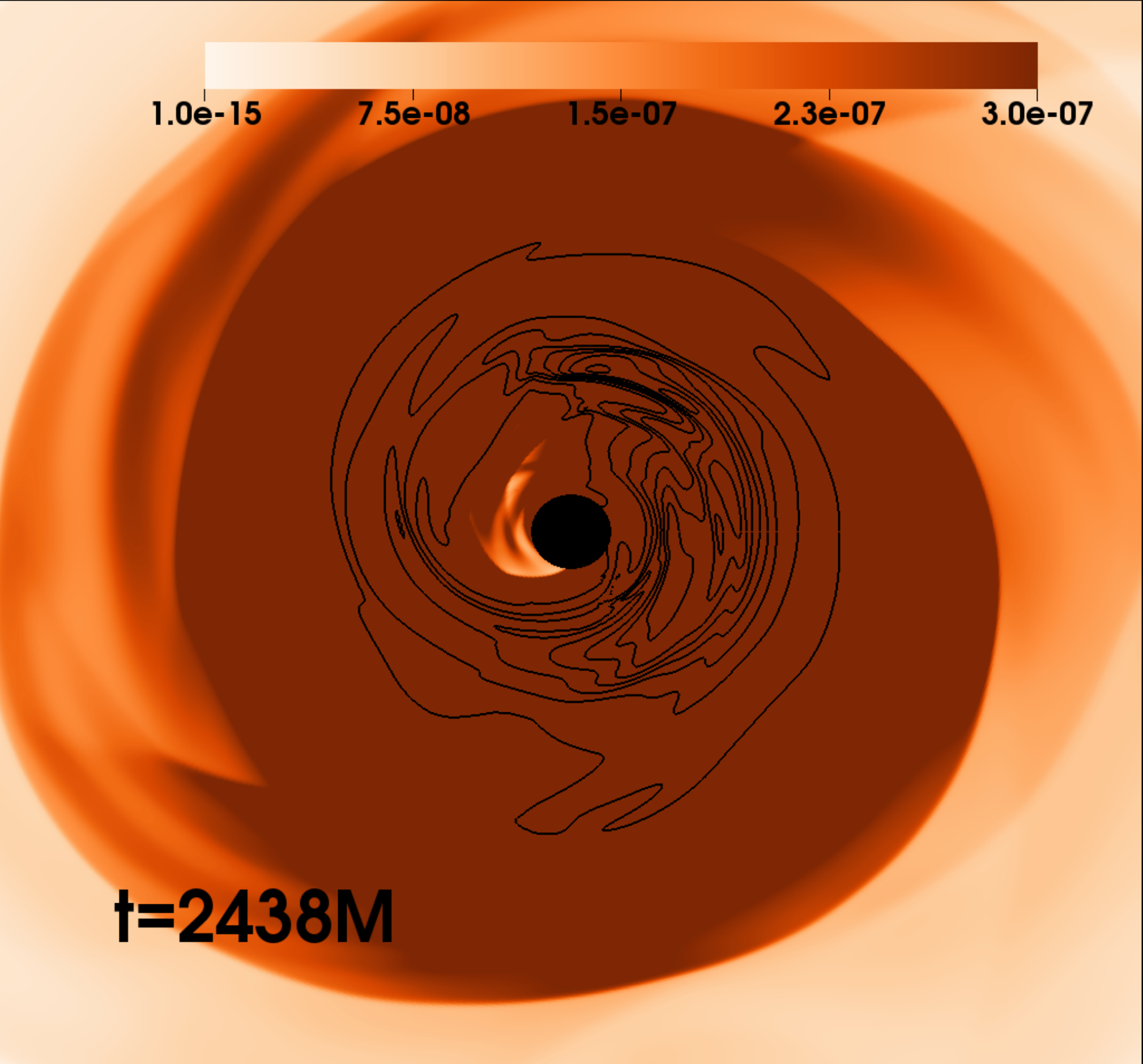}
 \includegraphics[width=0.31\textwidth]{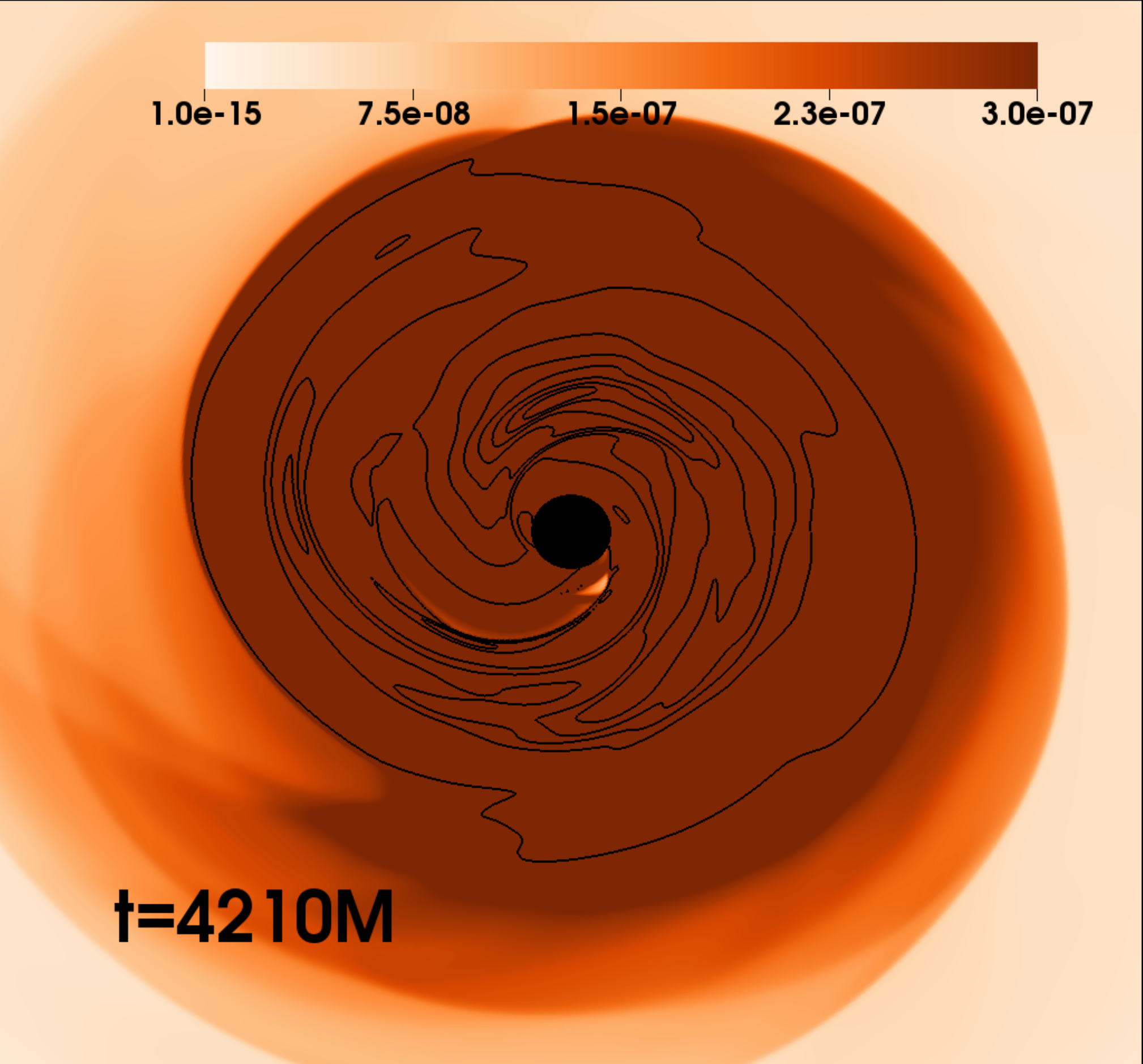}
 \includegraphics[width=0.31\textwidth]{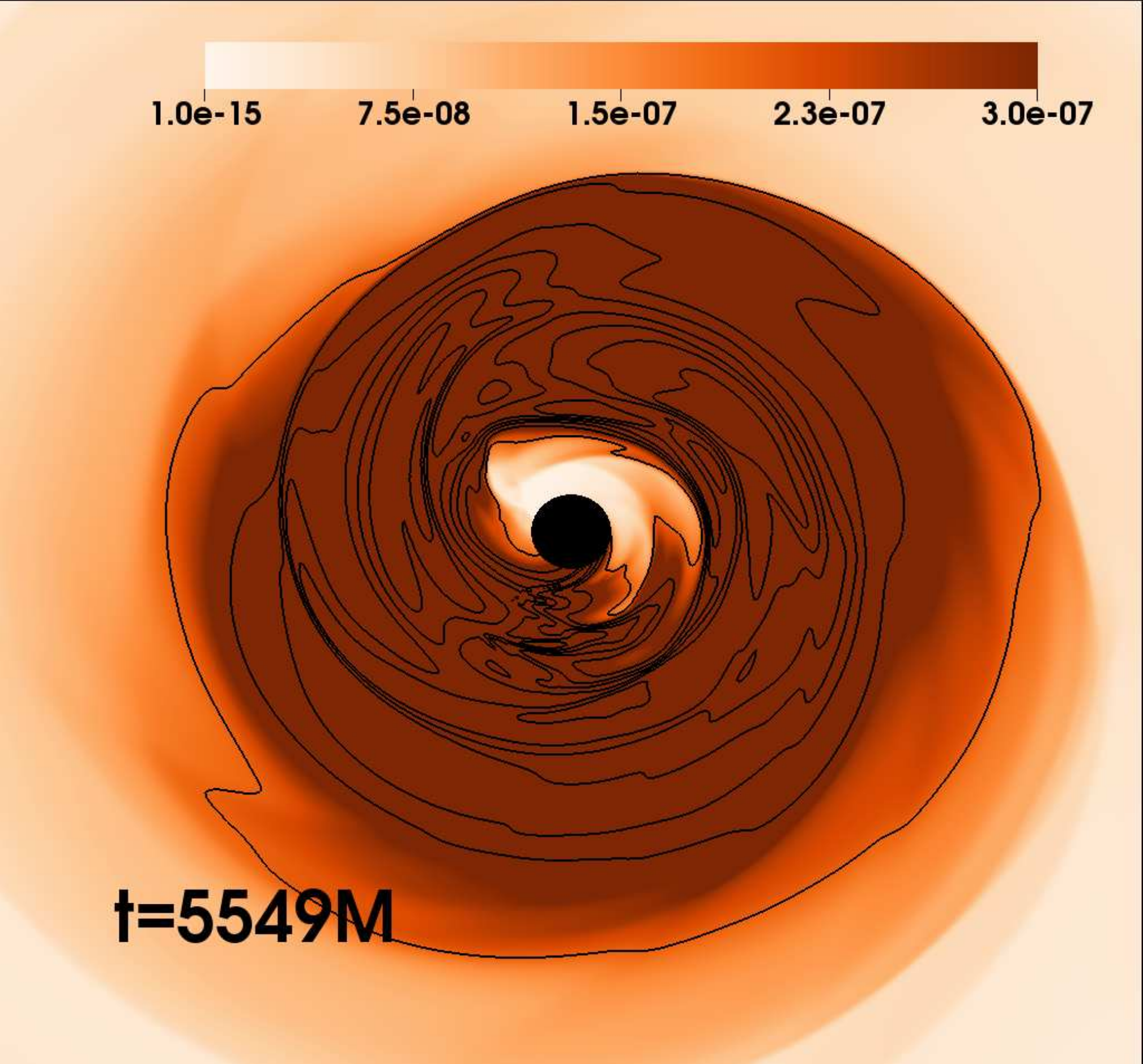}
 \includegraphics[width=0.31\textwidth]{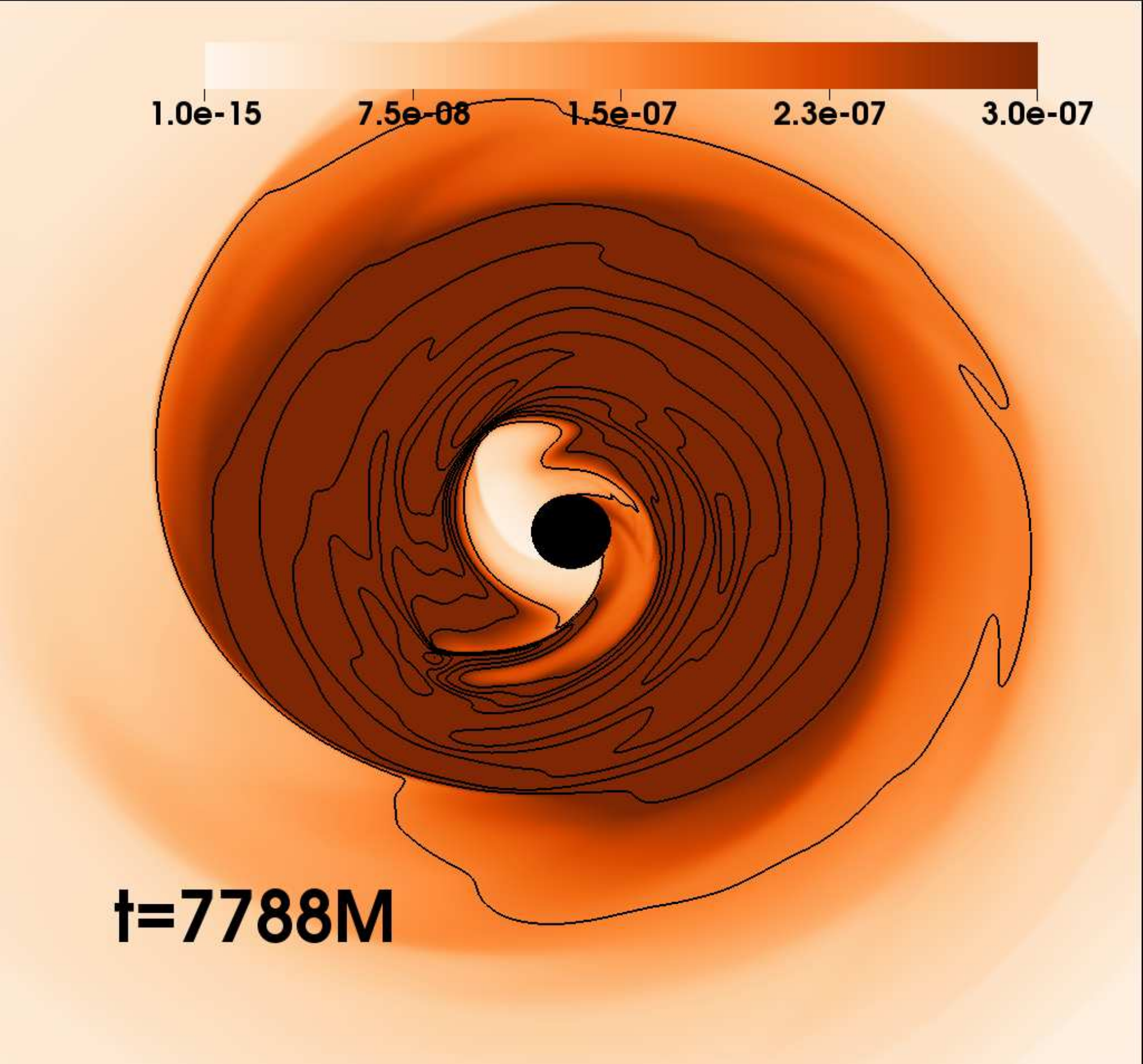}
 \includegraphics[width=0.31\textwidth]{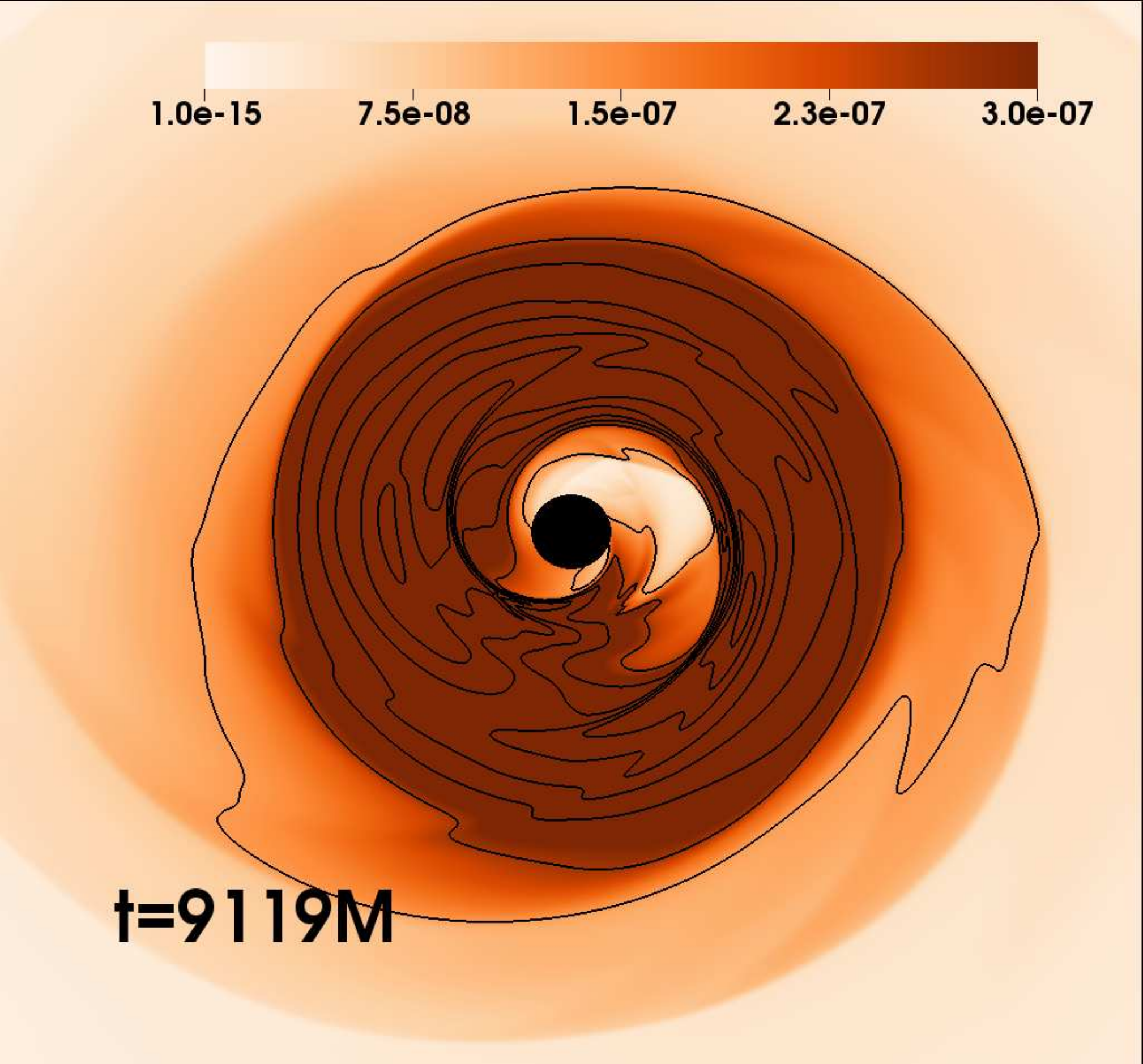}
 \includegraphics[width=0.31\textwidth]{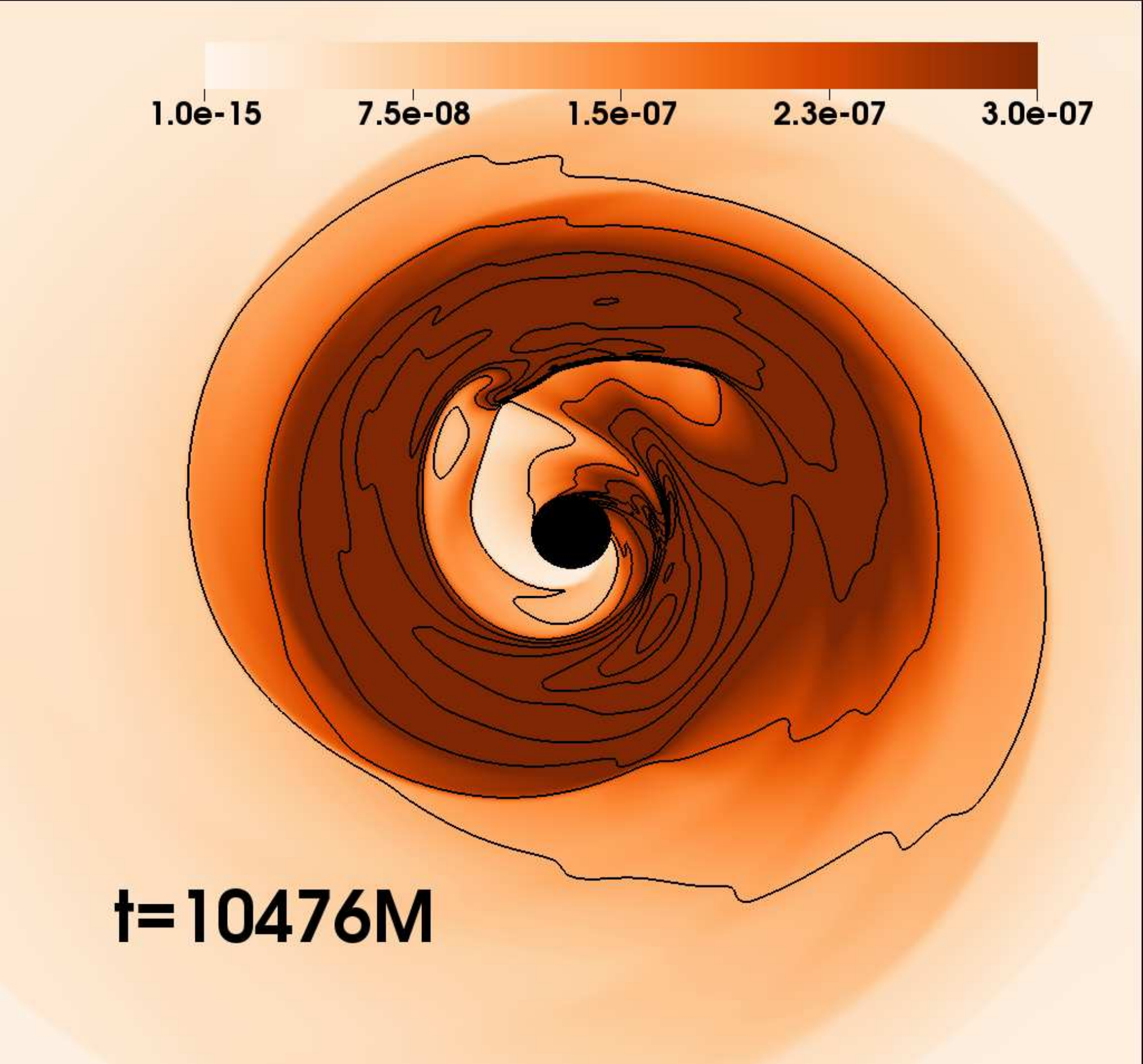}
 \includegraphics[width=0.31\textwidth]{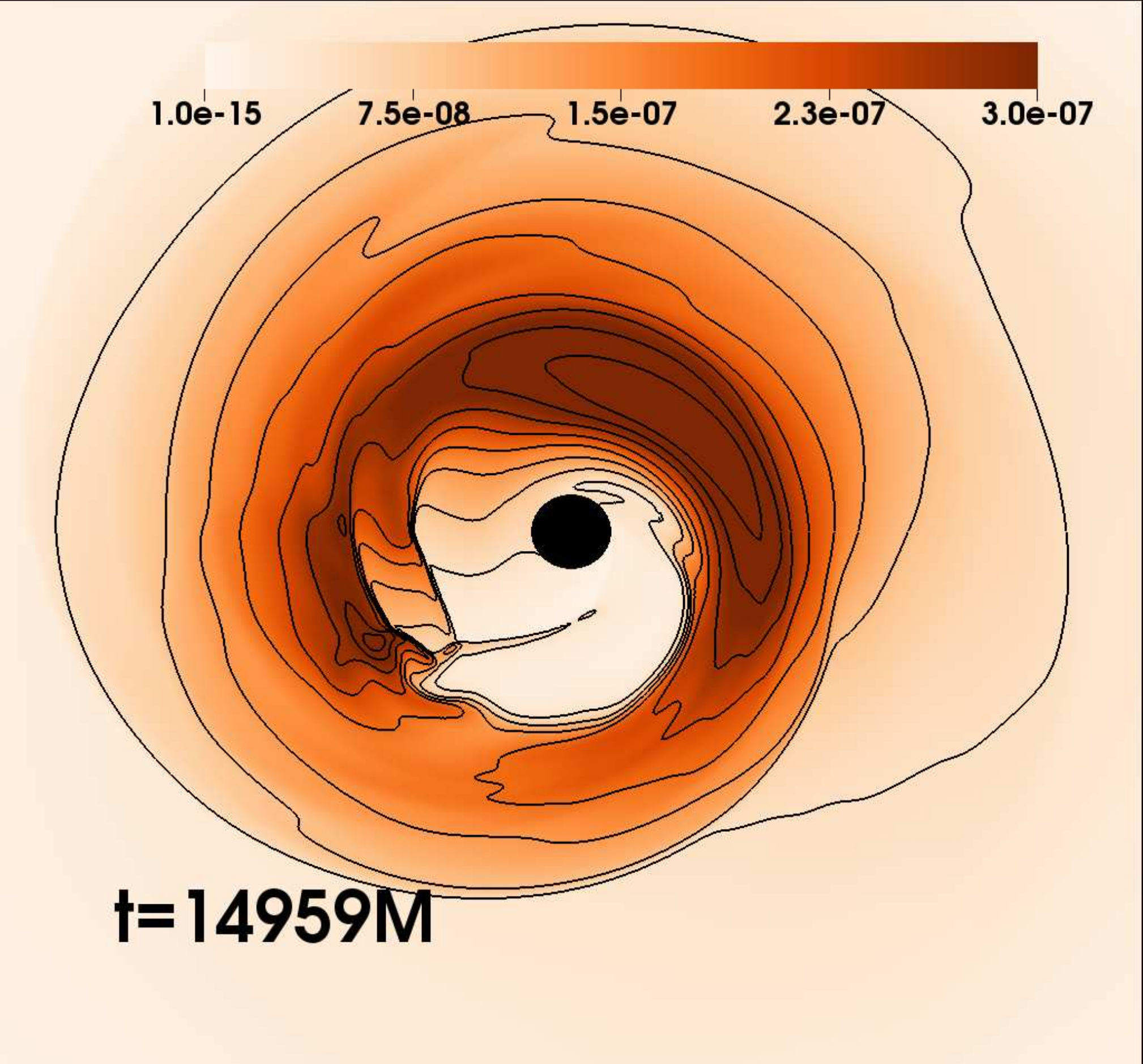}
 \includegraphics[width=0.31\textwidth]{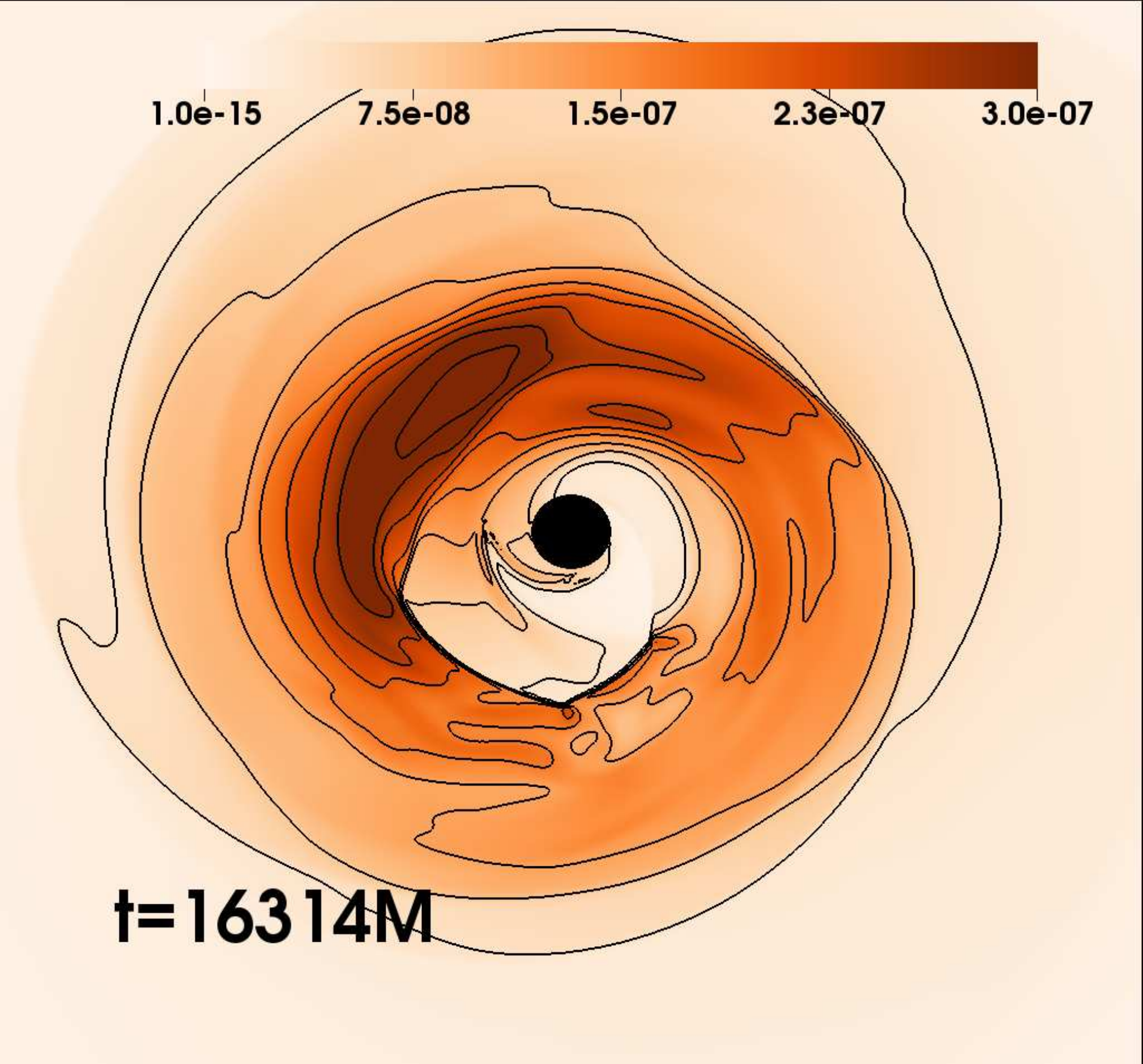}
 \includegraphics[width=0.31\textwidth]{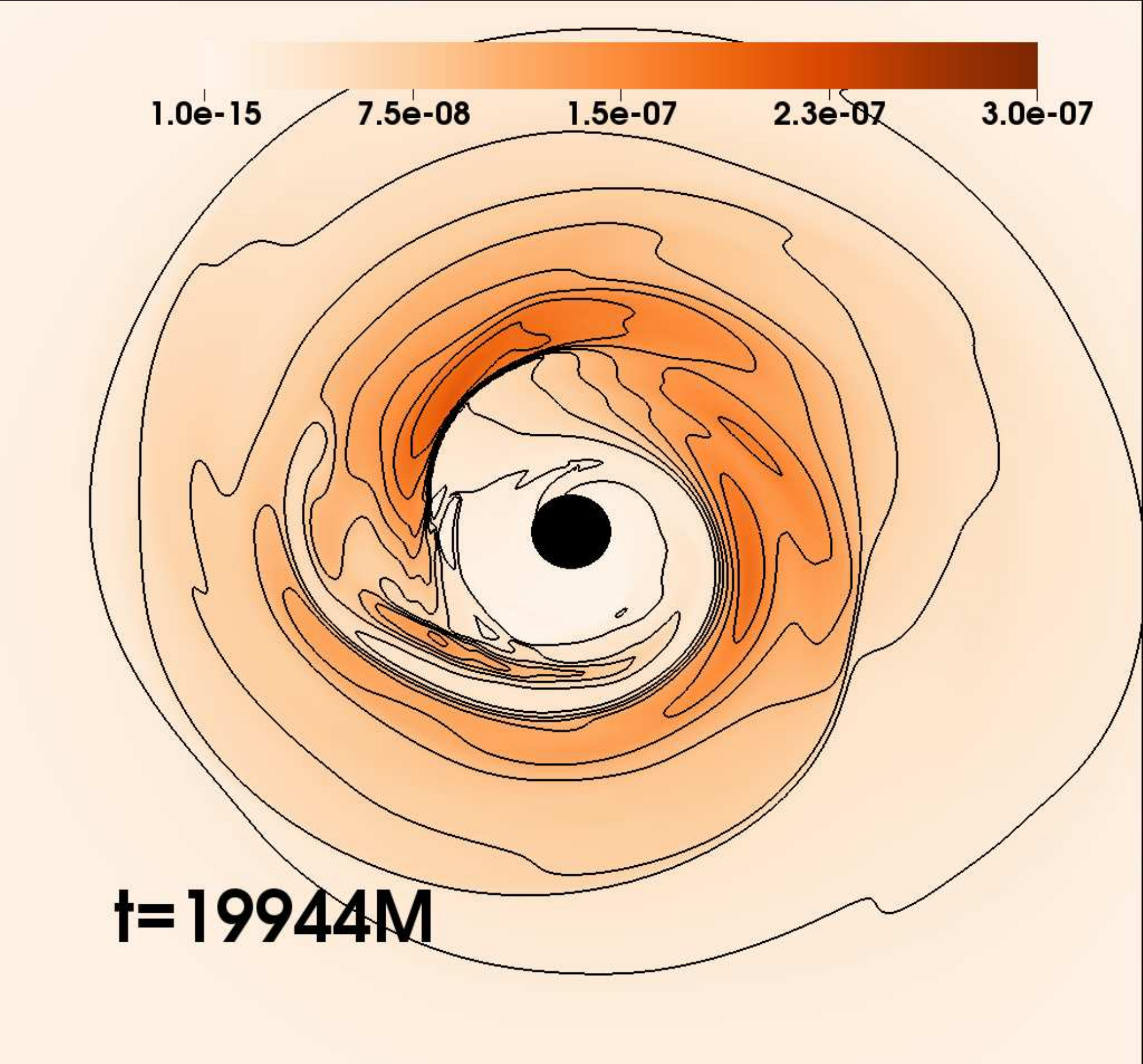}
 \includegraphics[width=0.31\textwidth]{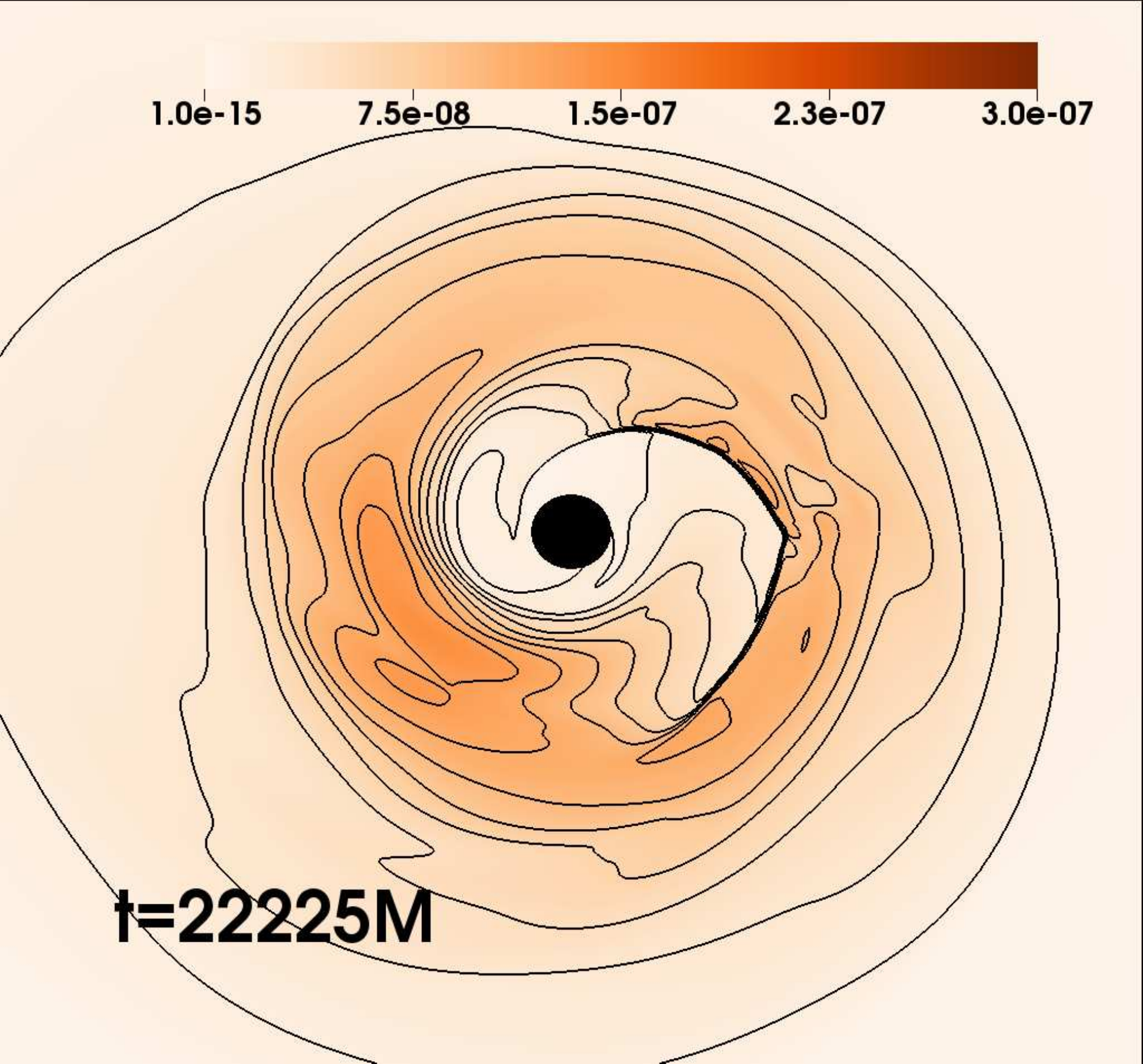} 
 \caption{The rest-mass density of the torus on
   equatorial plane with linearly spaced
   isocountours around the non-rotating black hole for the model $K00C$.
   {\bf The domain is $[X_{min},Y_{min}]
    \rightarrow [X_{max},Y_{max})] =[-40M,-40M)]\rightarrow [40M,40M]$.}}
\label{Non_rotating_1}
\end{figure*}

The dynamical evolutions of the rest-mass densities on the equatorial plane after the perturbation
show a big difference around the rotating and non-rotating black hole as seen in Figs.\ref{rotating_1}
and \ref{Non_rotating_1}, respectively.  The maximum rest mass density of the accreted
torus stays almost bellow its initial value
during the time evolution. Having a rotating or non-rotating black hole causes a drastic change on the dynamics
of the accreted matter and spiral waves created on the torus. 

\begin{figure}
 \center
\vspace{0.3cm}
\includegraphics[width=0.6\textwidth]{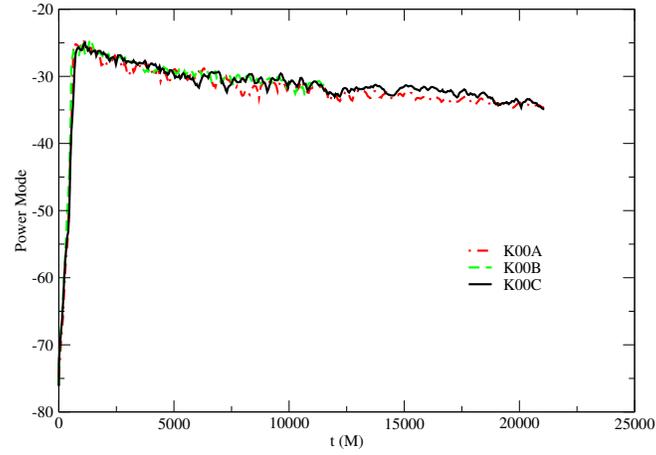}
\caption{The same as Fig.\ref{rotating1_5} but it is for the kicked torus around the
  non-rotating black holes.
\vspace{0.3cm}
\label{Non_rotating_3}}
\end{figure}

The kicked non-rotating black hole also leads some excitations of the specific unstable
non-axisymmetric modes. In order to compute the azimuthal number $m$
of the spiral wave mode, the dynamical evolution
of the torus is observed for different range of perturbation parameters
seen in Eq.\ref{perturb velocity}.
After the black hole-torus system is perturbed, the new
equilibrium  stage is almost reached at the saturation time $t=750M$, seen in
Fig.\ref{Non_rotating_3}. It shows that the kicked black hole
enhances the strength of the  fastest growing mode $m=1$.  spiral wave instability driven on the torus  can be
characterized by the presence of the
radial pressure gradient. As it is seen in Fig.\ref{rotating1_5}, the matter that is getting closer to the black hole
horizon would lead the stronger spiral density wave mode.

%
\subsection{Comparison of Rotating and Non-rotating Black Hole Cases}
\label{Compare}

\begin{figure}
 \center
\vspace{0.3cm}
\includegraphics[width=0.6\textwidth]{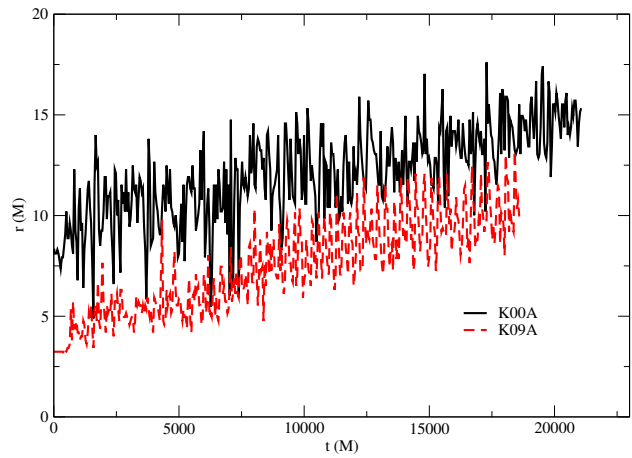}
\caption{The location of maximum value of density of the torus during the evolution
  around the non-rotating and rotating black holes. 
\vspace{0.3cm}
\label{compare 1}}
\end{figure}

The rest-mass densities for models at different snapshots are given around the
rotating and non-rotating black holes in Figs.\ref{rotating_1} and \ref{Non_rotating_1},
respectively. As it is seen from these two cases, the effect of black hole spin onto the torus as well as
the morphology of torus after perturbation are clearly seen.
As it is also seen in Fig.\ref{compare 1}, the matter is
pushed away from the center of the black hole
due to the oscillating black hole being connected to the black hole
horizon with a spiral wave, even
in the last snapshot of simulation for the rotating black hole this can be seen.
Similarly, we can also clearly see the structure of
the spiral arms close to the non-rotating black hole, shown in Fig.\ref{Non_rotating_1}.
We observed
significant differences in the morphology of the disk when the black hole is rotating
even though we used the same initial perturbation parameters.

\begin{figure}
 \center
\vspace{0.3cm}
\includegraphics[width=0.6\textwidth]{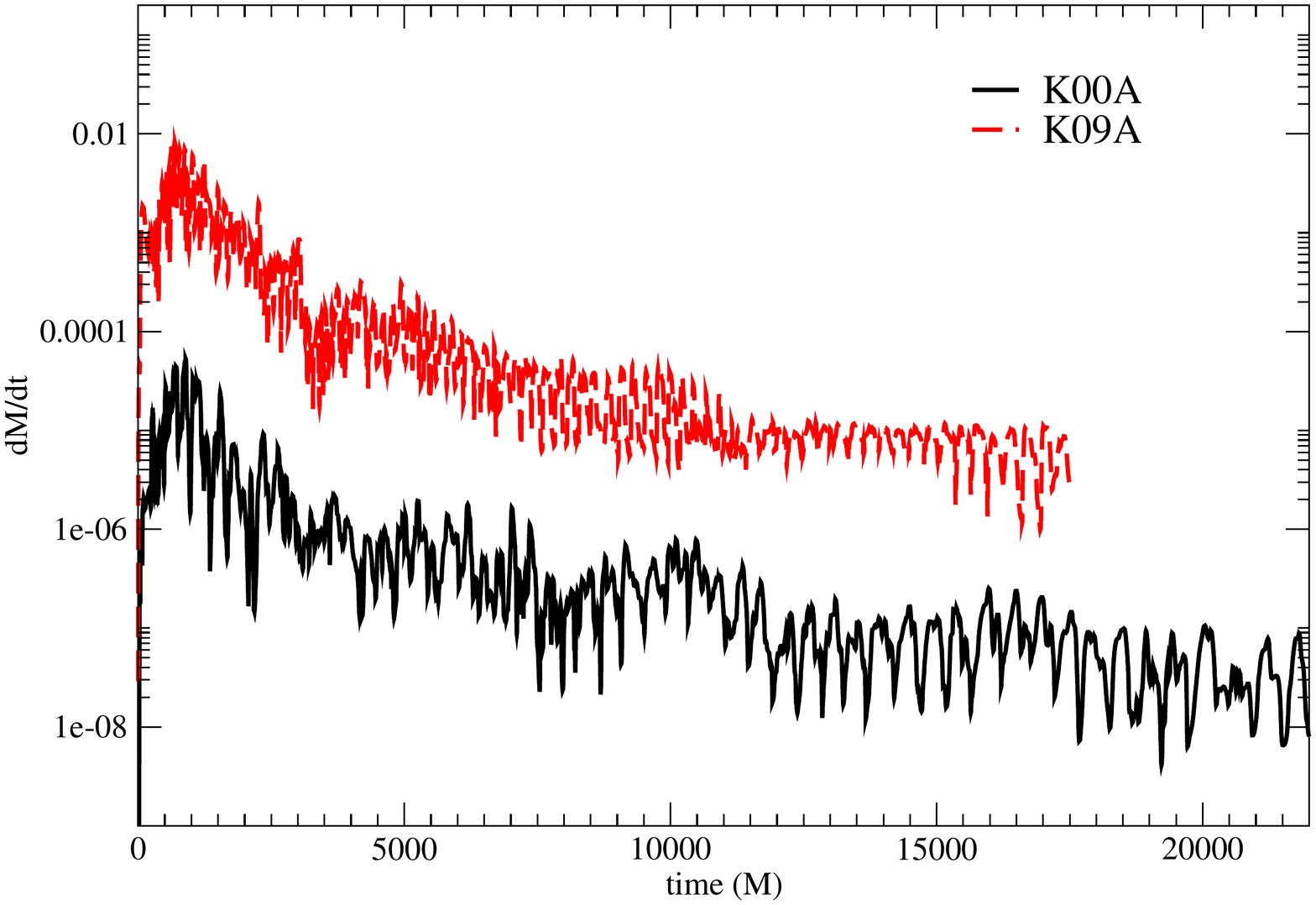}
\caption{Mass accretion rates for models computed in the inner boundary of
  the computational domain for models $K00A$ and $K09A$. 
\vspace{0.3cm}
\label{compare 2}}
\end{figure}

Overall, the mass accretion rates computed at the inner boundaries
around the non-rotating and the rotating black holes
show similar behavior during the evolution seen in
Figs.\ref{compare 1} and \ref{compare 2}, but
our numerical simulations of the perturbed torus due to the kicked black hole indicate that
transition from unstable state to steady-state shows clear evidence of dependencies on the
black hole spin. We find that the accretion rates are consistently bigger  for the larger black
hole spin. The reason for this is that the maximum density of the initial stable torus
around the rotating black hole
is closer to the black hole horizon. Kicked black hole makes a big impact on the torus dynamics
whenever the black hole is kicked.
 It is found that a higher spin produces a higher 
accretion rate, which implies the higher critical temperature, during the evolution. The increased
in the mass accretion rate causes the emitting radiation in the high-energy. Therefore, the torus-black hole
system may be a source of  continues high energetic emissions in several channels, possibly in association
with GRBs \citep{Putten3}.

The saturation time of the instability around the rotating  black hole is $t=1350$,
but it is almost half $t=750$ around the
non-rotating black hole. This means that the saturation times and growth
of the instability strongly depend on the black hole spin. After the saturations, the torus still keeps
losing the matter either towards or away from the black hole due to redistribution of the angular
momentum and we still witness the drastic changes in the torus dynamics during the evolution. Additionally,
as it is seen in Figs. \ref{rotating1_5} and \ref{Non_rotating_3}, the amplitude of the power of the $m=1$
mode around the rotating black hole is larger than that for the non-rotating black hole. It is a clear evidence for
dependency of the mode amplitude on the black hole spin.

Studying the angular momentum transfer around the black hole would tell us how the
accretion happens through the spiral shock wave.
In order to show the effect of the black hole spin on angular momentum transfer,
we compute the angular momentum flux at the location $r=6M$ inside the torus along the spherical surface.
The azimuthal component of the angular momentum flux in the equatorial plane is

\begin{eqnarray}
 \frac{dL}{dt} = -\int_0^{2\pi}\tilde{\alpha}\sqrt{\gamma}\rho h u^r u^{\phi} d\phi.
\label{GRH8}
\end{eqnarray}

\noindent Here, we will find out how the perturbed torus transports the angular momentum along
the equatorial plane; radially inward or outward and its dependencies to the black hole
spin parameter.

\begin{figure}
 \center
\vspace{0.3cm}
\includegraphics[width=0.6\textwidth]{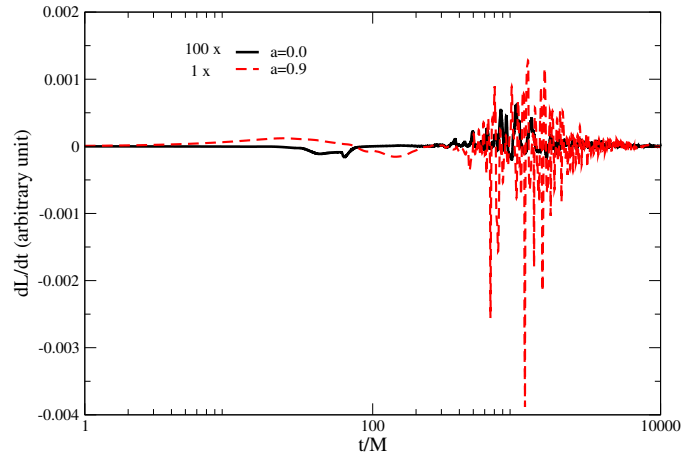}
\caption{The time evolution of the angular momentum flux distribution versus time
  at the location $r=6M$ around the rotating and non-rotating black holes.
\vspace{0.3cm}
\label{angularmomentum}}
\end{figure}

The perturbed torus around the black hole creates a spiral shock wave around the rotating
and non-rotating black holes. The angular momentum of the rotating gas can play an important
role in the moving of gas away from the black hole or towards it.
The angular momentum of the initial torus is constant in the begging
of simulation. After perturbation is felt by the torus, it is clearly seen in
Fig.\ref{angularmomentum} that the angular momentum flux varies depending on the black
hole rotation parameter. The strength of oscillating angular momentum flux is at least $100$ times
bigger in case of torus around the rotating black hole.
The higher the black hole spin parameter leads the bigger oscillation in the
angular momentum transfer. The angular momentum flux almost oscillates around
zero value for the rotating black hole. The spikes of the angular momentum transport
has a strong correlation with presence of spiral shock, seen in Fig.\ref{rotating_8}.

\begin{figure}
 \center
\vspace{0.3cm}
\includegraphics[width=0.6\textwidth]{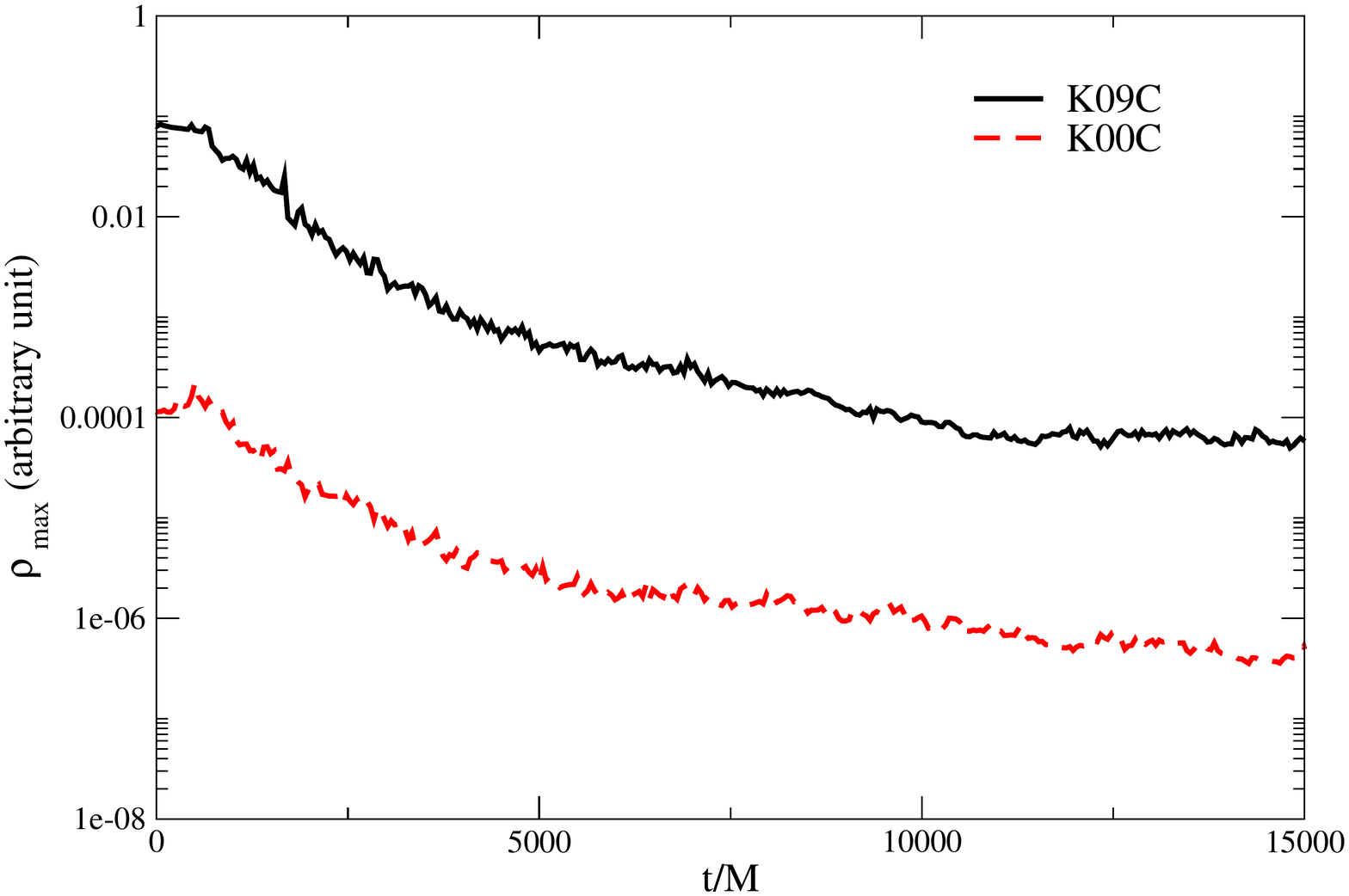}
\caption{The logarithmic maximum rest-mass density versus time for Models $K00C$ and $K09C$.
\vspace{0.3cm}
\label{perturbed_den}}
\end{figure}

In addition to power mode evolutions on the non-axisymmetric modes around the
rotating and non-rotating black hole given in Figs.\ref{rotating1_5} and  \ref{Non_rotating_3},
we explore more detail information about the local and non-axisymmetric
instability using the rest-mass density.  
The rest-mass density is one of the perturbed physical variable
of torus when the black hole oscillates inside the galaxy. The time evolutions
of the maximum value of
the rest-mass density in logarithmic scale around the non-rotating and rotating black holes
are given in
Fig.\ref{perturbed_den}. The non-axisymmetric perturbations display a significant
growth during certain time interval ( until $t\sim 10000M$) in both cases. It almost oscillates
around some equilibrium point after this time. The amplitude, however, strongly depends on the
black hole spin parameter.

%


\section{Discussion and Conclusion}
\label{Discussion and Conclusion}
We have studied the dynamical evolution of the torus around the black hole  that
undergoes  perturbation due to the kick, presumably it is the gravitational
radiation from the mergers of the black holes. There is a number of black hole-accretion
disk systems that might be considered as candidates for the  perturbed torus due to the kicked
black hole. The results found in our numerical simulations are given in geometrized units; therefore,
they might be used to explain many of the perturbed the kicked black hole-torus systems.

In order to understand the dynamical feature of the torus with respect to the perturbation amplitude
produced by the  kicked black hole, we have varied the parameter $\chi$, which is a
free parameter to control the perturbation on the torus radial velocity, and it is found that
different values of $\chi$ produce almost the  same behavior for given black hole-torus system.
Only some time delays when the spiral density wave instability was reached the saturation point and types of
shock waves created during the time  evolution were observed. It is clearly found that the instability seen for $m=1$
in the exponential growth mode is developed around the rotating  and non-rotating black holes.
The non-axisymmetric growing mode $m=1$ causes the formation of the spiral shock wave appeared in the rotating
matter around the black holes \citep{Mewes2}.

The mass accretion rates, which appear around rotating black hole almost
$100$ times bigger than the one  around the non-rotating black hole, 
show that the accretion associated with the saturation of spiral density growth mode
occurs slightly earlier in case of the rotating black holes.
Additionally, the location of the maximum density of the perturbed torus around the
rotating black hole gets closer with the non-rotating case after they reach
the saturation point, event though they are initially operated with a distance of $3.40M$.

We have observed the instability in our models which have developed due to the kick that
the center of black hole-torus system receives. The kicked black hole triggers the radial non-axisymmetric
oscillation on the disk. And these oscillations on the equilibrium stage of the torus lead to the spiral density
wave.
When the growing spiral azimuthal mode  occurs just before or after reaching of the saturation point could
trigger it not only in the torus, but also in the black hole. It would cause either
an increase or an decrease in the black hole kick velocity \citep{Mewes1}.

The rest-mass density oscillation, the growth of instability and the resulting spiral shock waves
that we have found in our
numerical simulations during the time evolution could play an important role in explaining the vigorous
X-rays phenomena observed from the galactical nuclei and  the quasars.

\section*{Acknowledgments}
The authors are grateful to the anonymous referee for constructive comments on the original manuscript.
All simulations were performed using the Phoenix  High
Performance Computing facility at the American University of the Middle East
(AUM), Kuwait.\\

\end{document}